\documentclass[compsoc,conference,a4paper,10pt,times,anonymous]{IEEEtran}
\IEEEoverridecommandlockouts

\usepackage{cite}
\usepackage{amsmath,amssymb,amsfonts}
\usepackage{algorithmic}
\usepackage{graphicx}
\usepackage{textcomp}
\usepackage{bmpsize}
\usepackage{lipsum}

\usepackage[colorlinks=true,citecolor=citecolor,urlcolor=black]{hyperref}
\usepackage{booktabs}
\usepackage{balance}
\usepackage{dblfloatfix}
\usepackage{fontawesome5}
\usepackage{multirow}
\usepackage[font=small,subrefformat=simple,labelformat=simple]{subfig}
\usepackage[absolute,showboxes]{textpos}
\usepackage{tikz}
\usepackage[table]{xcolor}
\definecolor{citecolor}{HTML}{BC1589}

\usetikzlibrary{
  arrows.meta,
  calc,
  colorbrewer,
  decorations.pathreplacing,
  fadings,
  fit,
  matrix,
  patterns,
  positioning,
  shapes.arrows,
}
\pgfdeclarelayer{bg}    %
\pgfdeclarelayer{fg}    %
\pgfsetlayers{bg,main,fg}  %

\graphicspath{{figs/}{figs/plots/}}

\usepackage{pifont}

\usepackage{xspace}
\newcommand{\eg}{\textit{e.g.,}~}
\newcommand{\ie}{\textit{i.e.,}~}
\newcommand{\cf}{\textit{cf.,}~}

\newcommand{\etal}{\textit{et al.}~}
\newcommand{\one}{({\em i})\xspace}
\newcommand{\two}{({\em ii})\xspace}
\newcommand{\three}{({\em iii})\xspace}

\usepackage{cleveref}
\crefname{appendix}{Appendix}{Appendices}
\crefname{equation}{Equation}{Equations}
\crefname{pluralequation}{Equations}{Equations}
\creflabelformat{equation}{#2\textup{#1}#3}
\crefname{section}{Section}{Sections}
\crefname{figure}{Figure}{Figures}
\crefname{table}{Table}{Tables}
\crefname{listing}{Listing}{Listings}

\makeatletter
\renewcommand{\paragraph}[1]{\noindent{\bf #1.}\hspace{0.25ex \@plus1ex \@minus.2ex}}
\makeatother

\makeatletter
\newcommand{\paragraphx}[1]{\noindent{\bf #1}\hspace{0.25ex \@plus1ex \@minus.2ex}}
\makeatother

\makeatletter
\newcommand{\paragraphS}[1]{\noindent{\bf #1?}\hspace{0.25ex \@plus1ex \@minus.2ex}}
\makeatother

\newlength\bitboxwidth%
\newlength\bitboxheight%

\begin{document}

\title{Secrets Best Not Shared: DNS Privacy Enhancements for the Constrained IoT}

\author{%
\IEEEauthorblockN{Martine S. Lenders}
\IEEEauthorblockA{\textit{TU Dresden}\\
Dresden, Germany \\
\href{mailto:martine.lenders@tu-dresden.de}{martine.lenders@tu-dresden.de}}
\and
\IEEEauthorblockN{Thomas C. Schmidt}
\IEEEauthorblockA{\textit{HAW Hamburg}\\
Hamburg, Germany \\
\href{mailto:t.schmidt@haw-hamburg.de}{t.schmidt@haw-hamburg.de}}
\and
\IEEEauthorblockN{Matthias W\"ahlisch}
\IEEEauthorblockA{\textit{TU Dresden} \& \textit{Barkhausen Institut}\\
Dresden, Germany \\
\href{mailto:m.waehlisch@tu-dresden.de}{m.waehlisch@tu-dresden.de}}
}

\maketitle

\definecolor{boxgray}{rgb}{0.93,0.93,0.93}
\textblockcolor{boxgray}
\setlength{\TPboxrulesize}{0.7pt}
\setlength{\TPHorizModule}{\paperwidth}
\setlength{\TPVertModule}{\paperheight}
\TPMargin{5pt}
\begin{textblock}{0.8}(0.1,0.04)
  \noindent
  \footnotesize
  If you refer to this paper, please cite the peer-reviewed publication: M.S. Lenders, T.C. Schmidt, M. Wählisch. 2026. Secrets Best Not Shared: DNS Privacy Enhancements for the Constrained IoT.
  In \emph{Proceedings of IEEE EuroS\&P 2026}. IEEE, Piscataway, NJ.
\end{textblock}

\bstctlcite{IEEEexample:BSTcontrol}

\begin{abstract}
Attackers often identify DNS traffic to disrupt or compromise Internet services.
While prior work has focused on encrypting queries using DNS over TLS, HTTPS, or QUIC to counter such attacks, we consider IETF protocols designed for resource-constrained IoT devices and empirically analyze the potential of obfuscating DNS traffic in addition to encryption.
We create a dataset of machine-to-machine-compatible data objects along with the corresponding DNS resolution processes, evaluating 296 deployment scenarios of resolving host names, including DNS over the Constrained Application Layer Protocol (CoAP) and an onion routing flavor of CoAP under varying link-layer conditions.
We compare them to DNS over HTTPS\@.
Using Random Forest and a header field analysis, we identify fields that leak most information.
Our findings show that DNS over CoAP with equalized packet lengths, block-wise transfer, and header compression reduces the accuracy of identifying DNS frames to 86\% and further to 77\% with payload compression.
Our approach outperforms DNS over HTTPS, where classifiers always identify DNS frames based on IP addresses.
The dataset is publicly available.
\end{abstract}


\section{Introduction}\label{sec:intro}

The Domain Name System (DNS) translates names to addresses.
Its original intent was to provide a ``human-readable dictionary'' of Internet hosts~\cite{RFC-1035}.
For the longest time, DNS exchanged this information without encryption.
This leads to privacy risks in many deployment scenarios because it leaks sensitive information~\cite{zhhwm-cdips-15,wz-dmdf-17,axrnf-kshpats-19,ppaa-ielsi-20,ccll-hsrdu-22}.
In addition, stateless DNS access exposes nodes too reflective, amplification attacks~\cite{r-ahrnp-14,msk-tbbft-17,hkccs-diorc-20,njsw-fsdat-21,ars-didss-23,kdsw-fhpmt-25}.
To mitigate both privacy and security risks, DNS over TLS~(DoT)~\cite{RFC-7858}, DNS over HTTPS~(DoH)~\cite{RFC-8484}, and DNS over QUIC~(DoQ)~\cite{RFC-9250} were introduced.

With the advent of the (low-end) Internet of Things~(IoT) it became apparent that established technologies require reconsideration. IoT devices are constrained in memory, CPU, and energy.
Most recently, we proposed DNS over the Constrained Application Protocol~(CoAP)~\cite{RFC-9953,lagns-snrid-23}, short DoC.
DoC, like DoH, was specifically designed to hide DNS traffic within regular application packets of CoAP (as HTTPS in DoH~\cite{cloudflare-dot-vs-doh}).

\begin{figure}
  \setlength{\belowcaptionskip}{-1.0em}
  \centering
  \includegraphics{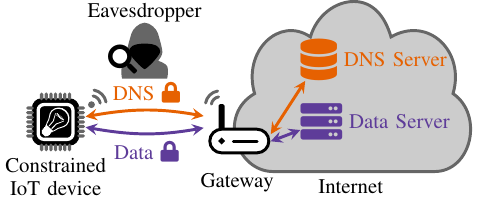}
  \caption{The attack scenario analyzed in this paper. An eavesdropper tries to distinguish data from DNS packets sent by a constrained IoT device on a wireless network.}%
  \label{fig:overview}
\end{figure}

Providing DNS security and privacy for the IoT is extremely important. 
First, DNS is a fundamental network service. 
An attacker who blocks DNS~traffic on the IoT may cause cascading outages because many IoT~services such as for smart cities or smart factories are interdependent.
Second, DNS leaks sensitive information that, for example, enables the identification of individual IoT devices~\cite{ppaa-ielsi-20,tmh-ridie-21}, which allows for targeted attacks.
To counter such attacks, it is crucial to prevent eavesdroppers from identifying DNS traffic, in particular in wireless domains, which ease wiretapping.
As soon as DNS traffic is identified, an attacker can interfere with DNS or apply DNS-specific analyzes.
Therefore, the scope of this paper are attackers connected via wireless networks to learn about traffic (see~\cref{fig:overview}) but insights are not limited to this scenario.

Encryption helps to improve security and privacy but does not solve the problem completely since an attacker can derive information from meta data.
Different from the core Internet, the constrained IoT deploys other means of packet modifications, which could help to hide DNS traffic in common IoT deployment scenarios~\cite{RFC-7452,RFC-8576}.
Static Context Header Compression (SCHC)~\cite{RFC-8724} was originally introduced by the IETF to allow for IP traffic in very constrained wide-area networks. %
The individual header compression scheme is only identified by an opaque rule ID but allows for the elision of headers.
The exchange of rules may happen out-of-band~\cite{RFC-9363,draft-toutain-schc-coreconf-management,draft-ietf-schc-universal-option,draft-ietf-core-comi}. Hence, SCHC offers a transparent opportunity to omit leaking header fields. CoAP provides block-wise transfer~\cite{RFC-7959}, which counters fragmentation and can ensure reliable transfer on the application layer. With these mechanisms in place, header data can be obfuscated, and packet lengths be equaled out with short block sizes.
Still, further compression techniques forced upon protocols may lead to even shorter DNS packets~\cite{draft-lenders-dns-cbor}, which in turn may  counter obfuscation.

In this paper, we question whether CoAP block-wise transfer combined with SCHC header compression significantly decreases distinction of encrypted DNS and data traffic when using DoC.
Additionally, we address aspects of recently proposed protocol extensions~\cite{draft-lenders-dns-cbor} that reduce the DNS packet sizes. %
To the best of our knowledge, this is the first paper that explores classification options of DoC within CoAP data traffic.
While we use established analyzes techniques set forth in related work, this work also provides detailed insights into basic OSCORE~\cite{RFC-8613} and Onion OSCORE~\cite{draft-amsuess-t2trg-onion-coap}.
Furthermore, we are the first to analyze the elision of header fields for privacy enhancements per peer-based SCHC rule sets.
In detail, our key contributions are:

\begin{enumerate}
  \item A comprehensive data corpus of HTTPS and CoAP traffic traces, combining data derived from the HTTP Archive~\cite{httparchive-bigquery} and corresponding DNS resource records (DoH and DoC) with 296 synthesized deployment options in unconstrained and constrained networks.
  This dataset does not only inform our study but may also serve as a basis for future DNS privacy research.
  We make this dataset publicly available.
  (\cref{sec:method-dataset,sec:data-corpus})
  \item A careful analysis of six machine learning (ML) classifiers to assess which ML classifier serves best for further exploitation of packet header fields.
  We confirm that Random Forest yields the best performance.
  (\cref{sec:method:analysis})
  \item A large scale analysis of header fields from the traffic traces using permutation feature importances of Random Forest to explore which header fields could leak DNS information to an attacker.
    (\cref{sec:eval:leaks})
  \item A proposal of peer-based SCHC rule sets to improve privacy by maximizing obfuscation of header information. These insights guide both more privacy-friendly DNS on the IoT and on the global Internet.
    (\cref{sec:eval:solutions})
\end{enumerate}

Prior work shows that DoT, DoH, and DoQ are still prone to information leakage when ML is used to distinguish DNS and data traffic~\cite{hlcw-iildot-19,cskd-pdohr-21,pammc-eiotp-24,hf-pldoq-24,mh-dcdoh-25,clzld-doqh3-25}.
They found that indicators such as inter-arrival times, header data, and packet lengths allow for fingerprinting DoH within HTTPS traffic~\cite{hlcw-iildot-19,cskd-pdohr-21,pammc-eiotp-24,mh-dcdoh-25}.
These metrics can even provide hints about the type of querying device or the requested resource. %
The constrained IoT with its low power, lossy link layers faces long delays and potential packet loss.
This renders the obfuscation of  inter-arrival times by adding artificial delays infeasible, as considered in prior work~\cite{pammc-eiotp-24}.
Similarly, the EDNS{(0)} padding option~\cite{RFC-7830} applied to obfuscate DNS packet lengths places incompatible burdens as it recommends to use multiples of 128~bytes for padding~\cite{RFC-8467}. This would inflate packet sizes and often enforce fragmentation, thus multiplying loss in the constrained IoT~\cite{lsw-ffln-21}, which should be avoided.
Following these constraints, this paper does not focus on expensive obfuscation techniques, but on the aforementioned block-wise transfer and header compression instead. Hence, we only look at packet formats.

The remainder of this paper is structured as follows.
In \cref{sec:background}, we provide background on the potentials of DoC and SCHC to hide DNS traffic.
\cref{sec:method-dataset} describes the methods we use to generate our data corpus of traffic traces and how we execute the header field analysis.
A quantitative analysis of the data corpus is presented in \cref{sec:data-corpus}.
In \cref{sec:method:analysis}, we introduce potential machine learning models and justify our selection of Random Forest for further analysis, followed by the detailed header field analysis and our proposal of possible advancements in \cref{sec:eval}.
We discuss related work in \cref{sec:related-work}, and limitations and our overall findings in \cref{sec:discussion}.
In \cref{sec:conclusion}, we conclude with an outlook on future work.

\section{Background}\label{sec:background}

In this section, we provide background on attack scenarios, the protocols, and principles we use and evaluate in this paper.

\paragraph{Attacks based on distinguishing DNS traffic}
Isolating encrypted DNS traffic can open a door to information leakage.
An attacker can use features such as the number, timing, or length of DNS messages to infer the type of websites being queried~\cite{hlcw-iildot-19}, the type of device making the query~\cite{pammc-eiotp-24}, or even classify the type of DNS traffic as desirable or undesirable~\cite{mh-dcdoh-25} in scenarios such as censorship.
Once isolated, an attacker can employ the very same machine learning techniques we are using in this paper.
More sophisticated machine learning techniques will only increase the accuracy of such an attack.
A key defense strategy is to make DNS traffic harder to detect within overall network traffic by obfuscating it through encryption, removing identifiable fields, and standardizing packet lengths using padding or block-wise transfers.
Obfuscating timing features is less of an option given the specific transmission times on the constrained IoT.

\paragraph{Obfuscating DNS with CoAP}
We proposed DNS over CoAP~(DoC)~\cite{RFC-9953,lagns-snrid-23} in the IETF to provide the benefits of DoH for the constrained IoT, \ie security and privacy of DNS traffic by encrypting and reusing an application layer protocol that is not DNS-specific.
CoAP~\cite{RFC-7252} is extended by two features.
\one Block-wise transfer to segment request and response bodies over UDP into equally-sized blocks~\cite{RFC-7959} and \two Object Security for Constrained RESTful Environments~(OSCORE)~\cite{RFC-8613} to provide end-to-end encryption on the application layer in addition to transport encryption based on DTLS~\cite{RFC-6347,RFC-9147}.
Block-wise transfer has the potential to homogenize length information of DoC and regular CoAP traffic. %
OSCORE adds a new mode of encryption that allows protecting header and payload information even from forwarding proxies.
To that end, Onion CoAP~\cite{draft-amsuess-t2trg-onion-coap} has been introduced, which utilizes multiple layers of OSCORE protection to provide Onion Routing~\cite{srg-pwb-97,dms-ttsgo-04} over several CoAP proxies.
In this paper, we refer to Onion CoAP as ``Onion OSCORE'' to distinguish it from {(D)}TLS-based Onion Routing.
Onion OSCORE has the potential of hiding the destination address and hostname of DoC messages from an observer and the source address from the DNS operator (\cf Oblivious DoH~\cite{semf-odns-19,schvv-odoh-21,RFC-9230,kdb-odoq-25}).

\begin{figure}
  \setlength{\belowcaptionskip}{-0.8em}
  \centering
  \includegraphics{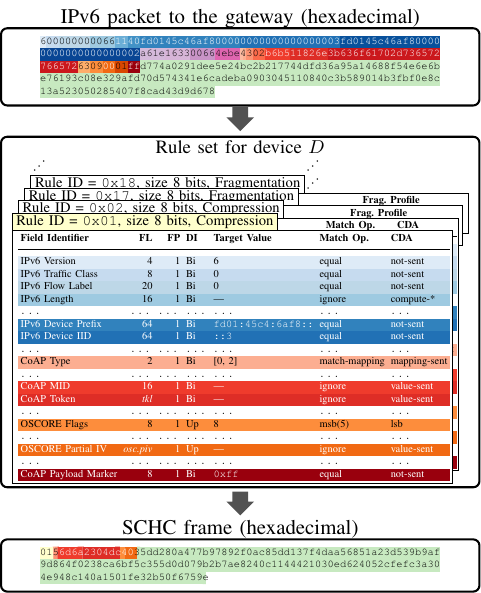}
  \caption{Static Context Header Compression (SCHC) uses rules (marked yellow) to compress and fragment IPv6 packets. Headers fields are identified by a \emph{field identifier} and are of a certain \emph{field length} (FL) at a \emph{field position} (FP).
  If they match for a certain \emph{direction} (DI) and a \emph{target value} using the \emph{match operator} (Match Op.), they compress to only a handful of bits (marked in shades of blue, purple, red, and orange) according to a \emph{compress/decompress action} (CDA). This may cause a bit shift (see, \eg payload marked in green). The ``not-sent'' and ``compute-*'' CDAs cause a header field to be completely elided from the SCHC frame.}%
  \label{fig:schc}
\end{figure}

\begin{figure*}
  \setlength{\belowcaptionskip}{-0.8em}
  \centering
  \input{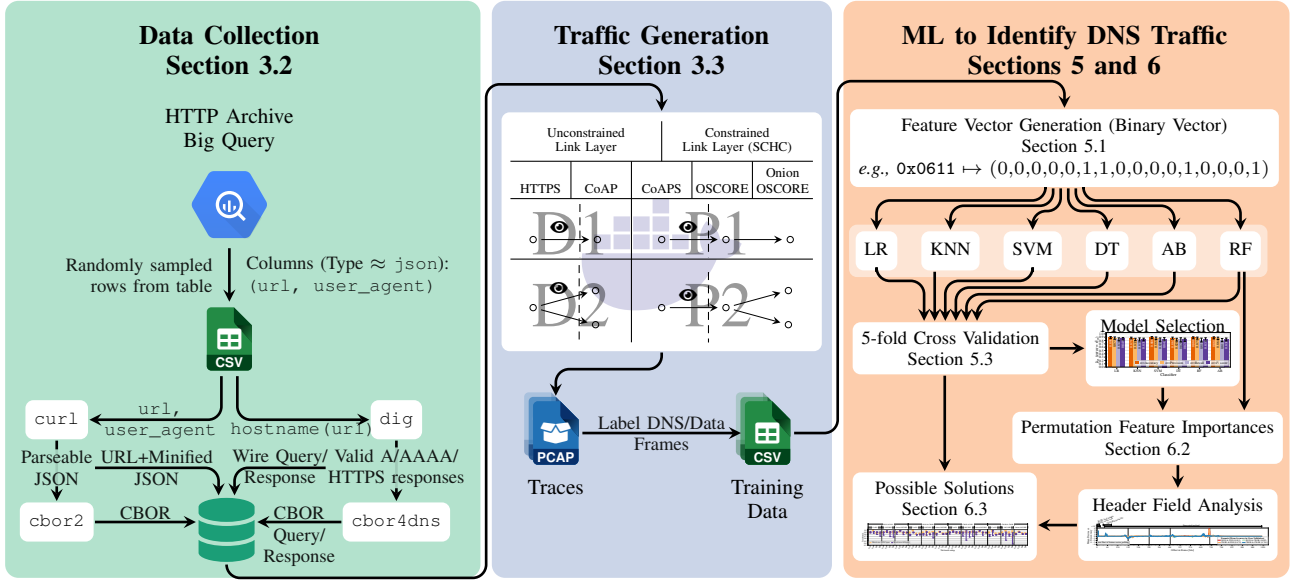}
  \caption{The method applied in this paper. We first randomly sample the HTTP Archive for data.
    Using that, we then generate traffic traces in a Docker-based environment.
    The feature vectors generated from the labeled traffic traces serve as input for cross validation of the ML models for model selection.
    We use RF to determine the permutation importances for every traffic scenario.
    These results in turn we use for our header field analysis to determine which header fields leak meta-information to an attacker.
    Last, we propose possible solutions based on our findings.}%
    \label{fig:method}
\end{figure*}

\paragraph{Eliding header fields}
Static Context Header Compression (SCHC)~\cite{RFC-8724} provides another option to obfuscate information in addition to encryption on the network, transport, and application layers.
SCHC is a framework for header compression originally designed for Low-Power Wide Area Networks (LPWANs)~\cite{RFC-8376}, but wider deployments, \eg for 6LoWPAN or interplanetary link layers are discussed~\cite{draft-ietf-6lo-schc-15dot4,draft-many-tiptop-ip-architecture}, highlighting the relevance of this protocol in future deployments.
SCHC uses rule sets---the static context---to define fragmentation and header compression of IPv6 packets on a per-device basis.
These rule sets are exchanged \emph{out-of-band}, \eg either directly burned into the firmware ROM or updated via an encrypted channel, \eg using CORECONF~\cite{draft-ietf-core-comi,draft-toutain-schc-coreconf-management,RFC-9363,draft-ietf-schc-universal-option}.
The rules in this rule set are identified by a rule~ID, which often, together with non-compressible residues, are the only information remaining in a packet after compression when sent from one Compressor/Decompressor (C/D) to another.
Other information is elided using the ``not-sent'' compression/decompression action (CDA), calculated implicitly using the ``compute-*'' CDA, or reduced to only a few bits, using CDAs such as ``mapping-sent'' or ``lsb'' (least significant bits), see~\cref{fig:schc}.

Since the best secrets are secrets that are not shared at all (\eg by eliding them), we see a potential for further obfuscation when using SCHC\@.
However, a few bits and even the rule ID can become an identifier for a message flow.
In this paper, we propose a \emph{peer-based} approach to SCHC rules to obfuscate flows between two peers.
The device ID that identifies rule sets usually is the link-layer address of the constrained device.
Combining source and destination address to the device ID associates rule sets with a pair of peers instead of just the device.
This way, we can use the same rule ID for multiple destinations and elide---rather than compress---flow information.

SCHC also provides a mechanism for fragmentation.
However, since this puts additional load on the link-layer, one typically wants to avoid this whenever possible~\cite{lsw-ffln-21}.

\paragraph{Using other DNS and data formats}
The media type \texttt{application/dns+cbor}~\cite{draft-lenders-dns-cbor} is a CBOR-based alternative to the common media type \texttt{application/dns-message}. %
Concise Binary Object Representation~(CBOR)~\cite{RFC-8949} is a binary alternative to JSON\@.
The aim of \texttt{application/dns+cbor} is to reduce DNS message sizes in constrained IoT deployments.
Reducing message sizes not only helps prevent fragmentation but (unfortunately) also makes it easier to identify DoC traffic within CoAP data, similar to how DoH can be recognized within HTTPS traffic~\cite{hlcw-iildot-19,cskd-pdohr-21,pammc-eiotp-24,mh-dcdoh-25}.

\section{Creating a Dataset for Privacy Analysis of DNS on the Internet of Things}\label{sec:method-dataset}

In this section, we describe our threat model and how we generate the traffic for our header field analysis.
The creation of our dataset is designed to gather real data where possible and combine this data with empirical measurements where real deployment is missing.
Therefore, we collect data objects based on Internet-wide measurements and extend them by synthesized traffic scenarios to reflect different IoT and DNS deployments.
\cref{fig:method} summarizes our method.

\subsection{Threat Model}\label{sec:method:threat}

Common IoT deployments connect constrained wireless devices to the Internet via a gateway~\cite{RFC-7452}.
The wireless domain poses a security challenge~\cite{RFC-8576}, as it creates an easy opportunity for pervasive monitoring attacks~\cite{RFC-7258}, \ie an adversary eavesdrops on the traffic of the IoT~device (see \cref{fig:overview}).

\paragraph{Target}
The target is the user or owner of an IoT device that the client is running on.

\paragraph{Adversary}
The adversary is an eavesdropper that collects the traffic between client and servers in a wireless domain, \eg using a radio device in promiscuous mode or a spectrum analyzer.
The adversary aims to differentiate between DNS and data traffic, \eg by fingerprinting the packet, in order to examine the DNS traffic in detail.

An attacker may use consumer grade devices to generate a labeled training set a~priori, similar to the approach we describe in this section.
Eavesdropper outside the wireless domain are not considered.
However, most obfuscation techniques discussed in this paper also apply to adversaries outside of the wireless domain. 
These include techniques based on header compression when the SCHC compressor/decompressor is not deployed on the gateway but closer to the servers and SCHC traffic is encapsulated~\cite{draft-ietf-schc-protocol-numbers}.

\paragraph{Threat}
Prior work has shown that identifying DNS traffic, even if encrypted, and extracting its communication patterns can lead to identification of device type, the category of sites requested by the device, or the type of DNS message exchange, see \cref{sec:background}.
Apart from these privacy concerns, this also poses security risks as device identification allows for exploits of known vulnerabilities.
Additionally, identifying DNS packets eases denial of service attacks since an attacker can disturb DNS packets on purpose to block communication.

\subsection{Data Collection}\label{sec:method:data}

Our data collection process is designed to reflect real data where possible to serve as a solid base for the evaluation of DNS privacy on the IoT.
Since the constrained IoT is emerging and DNS over CoAP is currently not widely deployed, we face the challenge to approximate future deployment.
The first step consists of identifying data objects that are fetched based on a hostname as part of a URL.
Those objects then require a DNS resolution.

\paragraph{Method}
Our dataset is based on HTTP Archive~\cite{httparchive-bigquery}.
From that database, we fetch the URLs of JSON objects, \ie REST-based, structured machine to machine (M2M) communication data.
For scalability reasons, we cannot consider all entries.
Instead, to collect responses, we randomly select entries with responses of the \texttt{summary\_requests} table\footnote{The schema of this database changed during our study. This information now resides in the \texttt{crawl} table, but the information is still there.}.
We select the user agent and URL to gain current information, \ie requesting real responses for a month from the actual servers using curl~\cite{curl}.
We discard \one all responses not parsable as JSON, and \two all responses larger than 1000 bytes in their minified JSON form, \ie when stripped of all non-essential whitespaces.
This ensures that our dataset is based on structured data (\ie M2M-capable) and aligns with requirements of the constrained IoT (small packet sizes).

To mix-in requests that have a payload---a typical CoAP traffic pattern---and to allow for block-wise requests, we also need data objects for the requests.
To that end, if a URL has more than 32 characters and has query parameters, we convert these parameters to an object.
Their keys become attributes to which we assign the value of the parameter.
If that object is larger than 1000 bytes as minified JSON, we discard it and the original response object to cap request sizes, too.
From those URLs, we remove the query parameters.

Then, we query the A, AAAA, and HTTPS DNS records of the URL hostnames using a public resolver
We collect HTTPS records since they also contain IP information~\cite{RFC-9490}. %
This reflects future deployment where DoC/CoAP-specific SVCB records~\cite{draft-ietf-core-transport-indication} guide DoC and CoAP requests.
To ensure repeatability and explainability of our results, we pre-collect these records and do not request them live during our traffic generation.

\paragraph{Setup}
In November 2024, we randomly picked 6,242,477~entries from the HTTP Archive.
Of those, we use the recorded pairs of the URL and user agent to request the real responses for a full month, February 2025.
This leads to 58,768~URLs with request and response objects.
We gain DNS data by resolving hostnames from a European vantage point via the public DNS~resolver~9.9.9.9.
We store the JSON object responses, including the URL, the converted query parameters, the HTTP status code that curl reported, and all successful DNS query and response messages.

We convert the collected JSON objects into CBOR using the \texttt{cbor2} library~\cite{cbor2} and convert the DNS messages to \texttt{application/dns+cbor} using \texttt{cbor4dns}~\cite{cbor4dns}, which implemented November 2024 version of CBOR~DNS~\cite{draft-lenders-dns-cbor-10} at the time of conversion.
When combining each data and DNS format, our measurements provide four~datasets of traffic generation, see \cref{tab:data-dns-formats}.

\begin{table}
  \caption{Data and DNS format dataset combinations.}%
  \label{tab:data-dns-formats}
  \scriptsize
  \centering
  \setlength{\tabcolsep}{2pt}
  \begin{tabular}{llr}
    \toprule
    \textbf{Data} & \textbf{DNS} & \textbf{Short Name} \\
    \midrule
    \texttt{application/json} & \texttt{application/dns-message} & JSON \& DNS\\
    \texttt{application/cbor} & \texttt{application/dns-message} & CBOR \& DNS\\
    \texttt{application/json} & \texttt{application/dns+cbor}    & JSON \& CBOR\\
    \texttt{application/cbor} & \texttt{application/dns+cbor}    & CBOR \& CBOR\\
    \bottomrule
  \end{tabular}
\end{table}

\subsection{Traffic Generation}\label{sec:method:network}

\begin{figure*}[h]
  \setlength{\belowcaptionskip}{-1em}
  \centering
  \subfloat[Direct, single server (D1).]{
    \centering
    \includegraphics{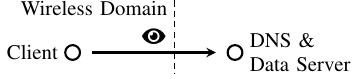}\label{fig:network-arch:d1}
  }\hspace{1em}
  \subfloat[Via proxy, single server (P1).]{
    \centering
    \includegraphics{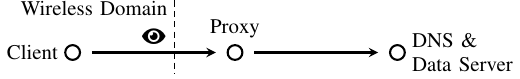}\label{fig:network-arch:p1}
  }\\
  \subfloat[Split server, direct (D2).]{
    \centering
    \includegraphics{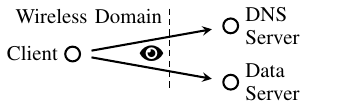}\label{fig:network-arch:d2}
  }\hspace{1em}
  \subfloat[Split server, via proxy (P2).]{
    \centering
    \includegraphics{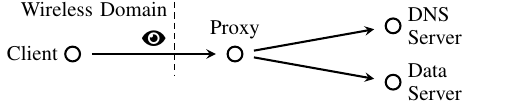}\label{fig:network-arch:p2}
  }
  \caption{Four setups how a client requests DNS and data. We use these for traffic generation. {\faIcon{eye}} marks the eavesdropper.}%
  \label{fig:network-arch}
\end{figure*}

Having a pre-collected set of data objects and principle DNS payloads at hand, we now need to create traffic traces that reflect how a client interacts with DNS and data servers.
These traces will serve as input to reproducibly emulate what an eavesdropper could observe and analyze.
This section describes the details of the different scenarios.

\paragraph{Network topologies}
We create traffic for four scenarios to provide a comprehensive view on common IoT deployments~\cite{RFC-7452,RFC-8576}
All scenarios are visualized in \cref{fig:network-arch}.
The \emph{client} represents a constrained IoT device, the \emph{dashed lines} a gateway, similar to the principle setup in \cref{fig:overview}.
Each proxy is located between a gateway and the servers to offer an additional level of obfuscation depending on the DNS transport.
When using SCHC, the upstream SCHC compressor/decompressor can either run on the gateway or between the gateway and the servers or the proxy.

\begin{enumerate}
  \item \emph{D1 (direct, single server)}, \cref{fig:network-arch:d1}: One shared server that serves both DNS and non-DNS data. The client communicates to directly with the server.
  \item \emph{P1 (via proxy, single server)}, \cref{fig:network-arch:p1}: One shared server that serves both DNS and non-DNS data. The client communicates to the server via proxy.
  \item \emph{D2 (direct, split server)}, \cref{fig:network-arch:d2}: Two servers that serve DNS and non-DNS data. The client communicates to both servers directly.
  \item \emph{P2 (split server, via proxy)}, \cref{fig:network-arch:p2}: Two servers that serve DNS and non-DNS data. The client communicates to both servers via a proxy.
\end{enumerate}

DoC was designed for constrained networks, \ie networks with low throughput, high packet loss, asymmetric links, and potential for fragmentation.
We assume these properties on the wireless link connecting the client to the Internet, monitored by the eavesdropper.

\paragraph{DNS protocols}
To compare different security modes of DoC with DoH\@,
we generate DoC traffic via DTLS (\ie \emph{CoAPS}) and with \emph{OSCORE} to secure messages.
Additionally, we use \emph{Onion OSCORE}~\cite{draft-amsuess-t2trg-onion-coap} with randomized CoAP MIDs and tokens.
For both DTLS and OSCORE, we use pre-shared keys~(PSKs) in AES128 encryption.
For baseline comparison, we run DoC using unencrypted \emph{CoAP}.

To test DoH, we rely on HTTP/2 for two reasons.
First, HTTP/2 is more prevalent than HTTP/3.
Second, using HTTP/2 instead of HTTP/3 allows us to gain insights into TCP effects on the classification of DNS and data traffic.

The DTLS and OSCORE libraries prefer AES128 PSKs, and the DTLS library only implements DTLS v1.2 at the time of evaluation.
To compare DoC and DoH fairly, we use TLS-PSK with TLS v1.2 and AES128 in DoH scenarios.
Both DoC and DoH servers, as well as the data servers, operate on their default port (443 for HTTPS, 5683 for CoAP and OSCORE, 5684 for CoAPS). %
For HTTPS, CoAPS, and Onion OSCORE, the proxy can obfuscate the intended target of requests and the origin of the server responses.
In fact, the effects of Onion OSCORE are only relevant when used in conjunction with a proxy, even in the P1 scenarios.

\paragraph{Details on resolving DNS and fetching data, and payload labeling}
We resolve the hostname carried in the URL by querying a DNS server.
We prioritize HTTPS over AAAA and AAAA over A records, depending on availability of data (see \cref{sec:method:data}).
For DoC, we use FETCH requests~\cite{RFC-9953}, for DoH POST\@.
The client does not cache the response, emulating IoT devices that query names at start-up, \eg for self-configuration, or after long sleep cycles when DNS response TTLs are expired~\cite{pammc-eiotp-24}.

After resolving a hostname, the client sends a request to the data server.
If a request object is available, the client serializes it to \emph{CBOR} or minified \emph{JSON}, depending on the scenario.
If no request object is available, the client uses~GET\@.

Including a real URL hostname in the unprotected part reduces the efficiency of SCHC rule definitions. %
To cover this, the client uses the placeholder hostnames ``dns-server'' and ``coap-server'' (or ``http-server'') in the CoAP \texttt{Uri-Host} option (or HTTP \texttt{Host} header).
These replacements \emph{only} happen in CoAP or HTTP headers to simplify SCHC rule deployment.
For the hostname resolution above, we still use the original URL hostname.
The server serializes the response object as \emph{CBOR} or minified \emph{JSON}, depending on the experiment.
It uses the HTTP Status Code or a corresponding CoAP response code, which we collected before.
We then strip the link-layer headers since this information can be randomized~\cite{hcb-romapp-22}.

We label the link-layer payloads as \emph{DNS frame} or \emph{data message}, depending on whether the fragment, segment, or block belongs to a DNS or data message.
For binary classification, we discard all frames not belonging to either category (TCP / {(D)}TLS handshake messages, empty CoAP ACKs,~\emph{etc.}).

\paragraph{Block-wise transfer}
Our payloads have a size of $\leq~1000$~bytes.
When evaluating block-wise transfer in CoAP, CoAPS, OSCORE, and Onion OSCORE scenarios, we specify a block size of 64~bytes.
For (Onion) OSCORE, we put the block options in the protected part of the message.

For non-block-wise scenarios, we leave the default block size value of the CoAP library (1024~bytes).

\paragraph{Link layer}
We send the packets either uncompressed and unfragmented, \ie over an \emph{unconstrained} link layer, or compressed and fragmented using \emph{SCHC} with a rule ID of length 8~bit, \ie simulating a 127~bytes PDU \emph{constrained} link layer.
In case of the unconstrained link layer, we use Ethernet.
For SCHC, we craft compression rules for each protocol and network architecture.
We only evaluate the CoAPS, OSCORE, and Onion OSCORE scenarios here.
For the split server scenarios (D2 and some P2), there are multiple possibilities to configure the SCHC compression rules:
\begin{enumerate}
  \item \emph{Split rules} (short \emph{split}): One rule for each server.
  \item \emph{Minimum rules} (short \emph{min.}): One rule for both servers and use the ``mapping-sent'' CDA to distinguish between the differences in the headers.
  \item \emph{Peer-based rules} (short \emph{peer}): One rule \emph{set} for each server.
  This uses a function of information from both peers as device ID\@. For our purposes, this function XORs the link-layer source and destination addresses of the peers and assumes both peers know these addresses of each other, \eg via CORECONF~\cite{draft-ietf-core-comi,draft-toutain-schc-coreconf-management,draft-ietf-schc-universal-option}.
\end{enumerate}

In the single server scenarios (D1 and P1, but also most P2), we use a single rule to compress data and DNS frames.
This elides and reduces header fields as much as possible. %
The protected part of OSCORE stays uncompressed~\cite{RFC-8824,draft-ietf-schc-8824-update} since OSCORE libraries rarely support this feature.

To compress CoAPS frames, we adapt a proposal by Fragkiadakis~\cite{f-dschci-22} to comply with our scenarios.
Control messages, such as empty ACKs or handshake messages, have their separate compression rules.

To implement fragmentation, we use No-ACK mode with datagram tag size 2, fragment compressed number field size 1, and reassembly check sequence (RCS) field size 4, using CRC32 for RCS\@.
Together with the 8-bit rule ID this results in a fragmentation header size of 11~bits for regular fragments and 15~bits for the final All-1 fragment.

\begin{table*}
  \caption{Overview of all scenarios for traffic generation. $Z$ is the block size. The numbers in each cell count the number of configurations per protocol as outlined in \cref{sec:method:network}.}%
  \label{tab:setup:traffic}
  \begin{center}
    \setlength{\tabcolsep}{2.5pt}
    \begin{tabular}{rr|cccc|cccccccc|cccccccc|cccccccc|cccccccc|l}  %
      \toprule
      \multicolumn{2}{r|}{\textbf{Formats}}                       & \multicolumn{36}{c|}{\bf Network Scenario} & \textbf{Sum} \\
                                                                          & & \multicolumn{20}{c|}{\bf Unconstrained Link Layer} &  \multicolumn{16}{c|}{\bf Constrained Link Layer (SCHC)} \\
                                                                          & & \multicolumn{20}{c|}{}                             &  \multicolumn{16}{c|}{\bf (3 variants if applicable)} \\
                                                                          & & \multicolumn{4}{c|}{\bf HTTPS} & \multicolumn{8}{c|}{\bf CoAP{(S)}} & \multicolumn{8}{c|}{\bf (Onion) OSCORE} & \multicolumn{8}{c|}{\bf CoAPS} & \multicolumn{8}{c|}{\bf (Onion) OSCORE}\\
                                                                          & & \multicolumn{4}{c|}{}         & \multicolumn{4}{c}{$Z = 1024$} & \multicolumn{4}{c|}{$Z = 64$} & \multicolumn{4}{c}{$Z = 1024$} & \multicolumn{4}{c|}{$Z = 64$} & \multicolumn{4}{c}{$Z = 1024$} & \multicolumn{4}{c|}{$Z = 64$} & \multicolumn{4}{c}{$Z = 1024$} & \multicolumn{4}{c|}{$Z = 64$} \\
                                                                          & & \multicolumn{2}{c}{\bf D} & \multicolumn{2}{c|}{\bf P} & \multicolumn{2}{c}{\bf D} & \multicolumn{2}{c}{\bf P} & \multicolumn{2}{c}{\bf D} & \multicolumn{2}{c|}{\bf P} & \multicolumn{2}{c}{\bf D} & \multicolumn{2}{c}{\bf P} & \multicolumn{2}{c}{\bf D} & \multicolumn{2}{c|}{\bf P} & \multicolumn{2}{c}{\bf D} & \multicolumn{2}{c}{\bf P} & \multicolumn{2}{c}{\bf D} & \multicolumn{2}{c|}{\bf P} & \multicolumn{2}{c}{\bf D} & \multicolumn{2}{c}{\bf P} & \multicolumn{2}{c}{\bf D} & \multicolumn{2}{c|}{\bf P} \\
      \multicolumn{1}{r}{\textbf{Data}} & \multicolumn{1}{r|}{\textbf{DNS}} & \bf 1 & \bf 2 & \bf 1 & \bf 2 & \bf 1 & \bf 2 & \bf 1 & \bf 2 & \bf 1 & \bf 2 & \bf 1 & \bf 2 & \bf 1 & \bf 2 & \bf 1 & \bf 2 & \bf 1 & \bf 2 & \bf 1 & \bf 2 & \bf 1 & \bf 2 & \bf 1 & \bf 2 & \bf 1 & \bf 2 & \bf 1 & \bf 2 & \bf 1 & \bf 2 & \bf 1 & \bf 2 & \bf 1 & \bf 2 & \bf 1 & \bf 2 \\
      \midrule
      \multirow{2}{*}{JSON}             & DNS                               & 1 & 1 & 1 & 1 & 2 & 2 & 2 & 2 & 2 & 2 & 2 & 2 & 2 & 2 & 2 & 2 & 2 & 2 & 2 & 2 & 1 & 3 & 1 & 1 & 1 & 3 & 1 & 1 & 2 & 6 & 2 & 3 & 2 & 6 & 2 & 3 & 74 \\
                                        & CBOR                              & 1 & 1 & 1 & 1 & 2 & 2 & 2 & 2 & 2 & 2 & 2 & 2 & 2 & 2 & 2 & 2 & 2 & 2 & 2 & 2 & 1 & 3 & 1 & 1 & 1 & 3 & 1 & 1 & 2 & 6 & 2 & 3 & 2 & 6 & 2 & 3 & 74 \\
      \multirow{2}{*}{CBOR}             & DNS                               & 1 & 1 & 1 & 1 & 2 & 2 & 2 & 2 & 2 & 2 & 2 & 2 & 2 & 2 & 2 & 2 & 2 & 2 & 2 & 2 & 1 & 3 & 1 & 1 & 1 & 3 & 1 & 1 & 2 & 6 & 2 & 3 & 2 & 6 & 2 & 3 & 74 \\
                                        & CBOR                              & 1 & 1 & 1 & 1 & 2 & 2 & 2 & 2 & 2 & 2 & 2 & 2 & 2 & 2 & 2 & 2 & 2 & 2 & 2 & 2 & 1 & 3 & 1 & 1 & 1 & 3 & 1 & 1 & 2 & 6 & 2 & 3 & 2 & 6 & 2 & 3 & 74 \\
      \midrule
      \multicolumn{2}{r|}{\textbf{Sum}}                                     & 4 & 4 & 4 & 4 & 8 & 8 & 8 & 8 & 8 & 8 & 8 & 8 & 8 & 8 & 8 & 8 & 8 & 8 & 8 & 8 & 4 &12 & 4 & 4 & 4 &12 & 4 & 4 & 8 &24 & 8 &12 & 8 &24 & 8 &12 &296 \\
      \bottomrule
    \end{tabular}
  \end{center}
\end{table*}

\paragraph{Total deployment scenarios}
Combining all possible configurations (\ie deploying all five protocols in all four network setups, considering two block sizes and all SCHC~rules), we create 296~deployment scenarios, see \cref{tab:setup:traffic}.
In the proxy scenarios of CoAPS over SCHC and in Onion OSCORE, server~identifiers are not visible in the unencrypted headers.
Therefore, the \emph{min.} and \emph{peer} rules apply only in D2 scenarios.
For plain OSCORE, the \emph{min.} scenario applies in P2 scenarios because the \texttt{Uri-Host} option, which contains the server name, is unprotected.
Consequently, we have the following number of variations.
In CoAPS, 3 variants for each D2 scenario (all 3 SCHC variants), in OSCORE 6 variants for each D2 scenario (both OSCORE variants and all 3 SCHC variants) and 3 for each P2 scenario (both OSCORE variants but 2 SCHC variants only for plain OSCORE).

\section{Overview of Data Corpus}\label{sec:data-corpus}

\begin{figure}
  \setlength{\belowcaptionskip}{-1em}
  \subfloat[Minified JSON objecs.]{
    \centering
    \includegraphics{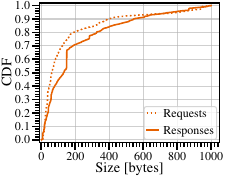}\label{fig:data-generation:json}
  }\hfill
  \subfloat[Classic DNS messages.]{
    \centering
    \includegraphics{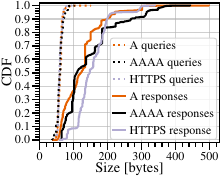}\label{fig:data-generation:dns}
  }\\[-.5em]
  \subfloat[CBOR objects.]{
    \centering
    \includegraphics{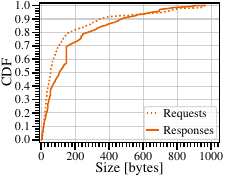}\label{fig:data-generation:cbor}
  }\hfill
  \subfloat[DNS CBOR messages.]{
    \centering
    \includegraphics{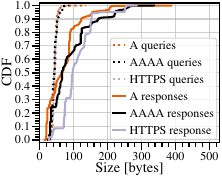}\label{fig:data-generation:cbor-dns}
  }
  \caption{Size distributions of differently encoded data objects and DNS~messages. Note that $x$-axis scaling differs for data objects and DNS messages.}%
  \label{fig:data-generation}
\end{figure}

In this section, we provide an overview of the data corpus that we created (see \cref{sec:method-dataset}), which we make publicly available.
Our set of data objects contains 58{,}768 request and response objects each.
8{,}541~URL hostnames for these object pairs resolve to HTTPS records.
18{,}543 hostnames resolve to AAAA records.
10{,}691 hostnames have an AAAA records but no HTTPS record and 39{,}533 hostnames exclusively resolve to A records.
All hostnames with an HTTPS record also have a corresponding AAAA record and all names with an AAAA record also have an A record.

\cref{fig:data-generation} shows the CDFs for the data objects and DNS message sizes.
80\% of JSON request objects are 189 bytes long or shorter, while 80\% of CBOR request objects are 165 bytes long or shorter.
80\% of JSON response objects are 326 bytes long or shorter, while 80\% of CBOR response objects are 274 bytes long or shorter.
99\% of classic DNS queries are smaller than 83 bytes and 150 bytes max.
The responses can be up to 498 bytes long.
CBOR-based DNS queries are a maximum of 136 bytes long.
Their responses can be up to 388 bytes long.
Both CBOR-based message types are shorter than classic DNS or JSON\@.
DNS messages encoded in CBOR are also much shorter than even CBOR-serialized data objects.

\begin{figure}
  \setlength{\belowcaptionskip}{-1em}
  \centering
  \includegraphics{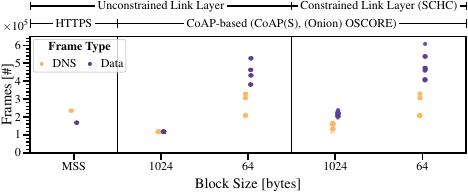}
  \caption{The number of frames per scenario, grouped by protocol and link layer (top), and block size (bottom). In HTTPS, we use the Maximum Segment Size.}%
  \label{fig:dns_data_frames_scatter}
\end{figure}

The number of frames generated in each scenario is shown in \cref{fig:dns_data_frames_scatter}.
In the scatter plot, each data point represents one scenario.
They are grouped by protocol and link layer (top $x$-axis), and block size configuration (bottom $x$-axis).
The HTTP/2 library splits headers and payloads of each stream into two separate TCP exchanges, causing HTTPS to have approximately double the number of frames than unconstrained CoAP with block size 1024 bytes in all security modes.
Naturally, more frames are generated with SCHC fragmentation and CoAP with smaller block sizes.

\begin{figure*}
  \setlength{\belowcaptionskip}{-1em}
  \setlength{\abovecaptionskip}{0.5em}
  \centering%
  \subfloat[Mean lengths]{%
    \centering%
    \includegraphics{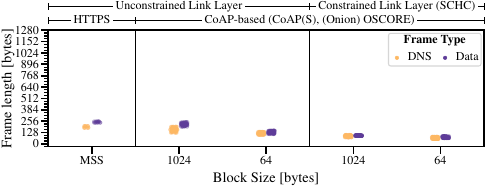}\label{fig:frame_lengths_scatter:mean}%
  }\hfill%
  \subfloat[Minimum and maximum lengths]{%
    \centering%
    \includegraphics{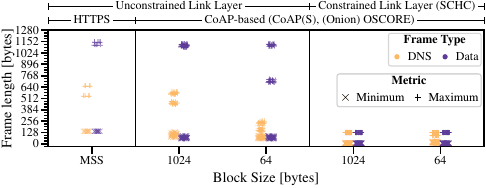}\label{fig:frame_lengths_scatter:min_max}%
  }%
  \caption{The mean, minimum, and maximum lengths of frames per scenario, grouped by protocol and link layer (top $x$-axis), and block size (bottom $x$-axis). In HTTPS, we use the Maximum Segment Size.}%
  \label{fig:frame_lengths_scatter}
\end{figure*}

The lengths of the frames generated in each scenario are shown in \cref{fig:frame_lengths_scatter} as scatter plots similar to \cref{fig:dns_data_frames_scatter}, separated by mean (\cref{fig:frame_lengths_scatter:mean}) as well as minimum and maximum (\cref{fig:frame_lengths_scatter:min_max}).
In general, DNS frames are shorter than data frames, due to their payload lengths (see \cref{fig:data-generation}).
Even with block size 64~bytes, data frames reach sizes of over 1 kByte in unconstrained scenarios.
Unusually long URI paths in the CoAP header are the cause of these larger frame sizes.
Only the body is split into blocks, options in unprotected headers are not segmented.
Undesired SCHC fragmentation \emph{together} with a block size of 64~bytes leads to similar length distributions between DNS and data frames.

\section{Machine Learning to Identify DNS Traffic}\label{sec:method:analysis}

We use established machine learning techniques and supervised learning algorithms to create an algorithmic foundation for our header field analysis.
In the following, we discuss all necessary steps, which are also visualized in \cref{fig:method}.

\begin{figure}
  \setlength{\belowcaptionskip}{-0.5em}
  \centering
  \includegraphics{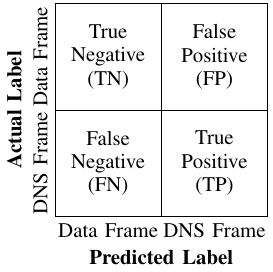}
  \caption{Confusion matrix of our binary traffic classification.
  Data frames are frames that belong to a data request or response,
 DNS frames are frames that belong to a DNS request or response.}%
  \label{fig:confusion_matrix}
\end{figure}

\subsection{Feature Vector Generation}\label{sec:method:analysis:vec}
First, we generate the feature vector for 5-fold cross validation and permutation feature importances.
We only regard packet formatting because prior observations render obfuscation techniques to manipulate packet timings infeasible due to restrictions put onto the sender by the constrained link~\cite{pammc-eiotp-24}.
Due to our emulated network scenarios, timings also do not yield realistic observations.
Our goal, after all, is to algorithmically determine leaking header field information.
Since our dataset is very heterogeneous, the entire packet is needed for cross validation and determining feature importances.
SCHC operates on the bit level.
Therefore, we translate the bytes of the observed Ethernet payloads and SCHC frames into \emph{binary vectors}, \ie each byte $B_i$ is translated to an 8-dimensional vector, where the bits $b_{i, j}$ of $B_i$ are the components of the vector.

\begin{equation*}
  B_i = \sum_{j = 0}^{7} b_{i, j} 2^j \mapsto \left( b_{i, j} \right)_{j = 0}^{7} \qquad b_{i, j} \in \{0, 1\} \label{eq:analysis:byte2vec}  %
\end{equation*}

We concatenate the resulting array of vectors to one large vector.
\begin{align*}
  \left[ B_i \right]_{i = 0}^{n} &= \left[ \left(b_{i,j} \right)_{j = 0}^{7} \right]_{i = 0}^{n} \\  %
  \left[ B_i \right]_{i = 0}^{n} &\mapsto (b_{0,1}{,}{\ldots}{,}b_{0,7}{,}{\ldots}{,} b_{n,1}{,}{\ldots}{,} b_{n,7})  %
\end{align*}

Input vectors need to be real number vectors of the same length to be processed by classification models.
We pad vectors of smaller dimension with ``2''.
Since all vectors used to be packets at byte level, the number of 2s added to each vector is a multiple of 8.

\begin{equation*}
  (b_{0,1}{,}{\ldots}{,} b_{n,7}) \mapsto (b_{0,1}{,}{\ldots}{,} b_{n,7}{,}{\ldots}{,}2{,}2{,}2{,}2{,}2{,}2{,}2{,}2)
\end{equation*}

We then use Min-Max-Normalization~\cite{jnr-snmbs-05} to scale the components of all vectors $X$ to a range between 0 and 1, \ie 0 remains 0, 1 becomes 0.5, and 2 becomes 1.
\begin{align*}
  X_{\text{scaled}} = \frac{X - \min(X)}{\max(X) - \min(X)}
\end{align*}

\subsection{Possible Machine Learning Models}
We base our analysis on common classifiers used in related work~\cite{hlcw-iildot-19,cskd-pdohr-21,aaazs-rsnto-23,pammc-eiotp-24,alabk-tbuid-25}, namely Logistic Regression (\emph{LR}), $k$-Nearest Neighbors (\emph{KNN}), Support Vector Machine (\emph{SVM}), Decision Tree (\emph{DT}), Random Forest (\emph{RF}), and AdaBoost (\emph{AB}).
We also considered Na\"{\i}ve Bayes~(\emph{NB}), but initial tests showed underwhelming performance in our scenarios.
We explicitly do not look into deep learning classifiers, as most prior work on DoH only used classic feature-based classifiers as well~\cite{cskd-pdohr-21,pammc-eiotp-24,mh-dcdoh-25} and we want to \one stay comparable with these prior work and \two use the features to be able to interpret and explain leaking information.
To compare the models, we use 5-fold cross validation with \emph{scikit-learn}~\cite{pvgmt-slmlp-11}, or---for all but DT and AB---the CUDA-based alternative \emph{cuML}~\cite{rpn-mlpmd-20}.
We use most default parameters of \emph{scikit-learn}/\emph{cuML}, \ie no fixed random state and 5 unshuffled, stratified folds for 58,768 DNS and data message pairs, each, on the application layer---even more when considering fragmentation and block-wise transfer.
For LR, we perform a maximum of 5000~iterations, for KNN we use brute-force search to compute the nearest neighbors, for SVM we use $C = 0.01$, for RF we use 250 trees with a maximum depth of 9 in the forest, and for AB we use 250 estimators.
These numbers represent a rough average of the values used in related work~\cite{hlcw-iildot-19,cskd-pdohr-21,aaazs-rsnto-23,pammc-eiotp-24,alabk-tbuid-25}.

\begin{figure}
  \setlength{\belowcaptionskip}{-1em}
  \centering%
\subfloat[Mean of means $\bar\mu_m$ for $m \in$ \{Accuracy, Precision, Recall, $F_1$\}.]{%
    \centering
    \includegraphics{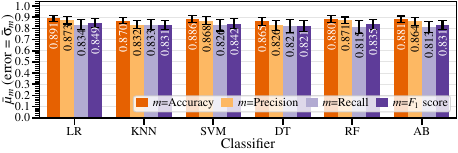}\label{fig:model_selection:metrics}%
  }\\[-0.1em]%
  \subfloat[Fit \& score times.]{%
    \centering
    \includegraphics{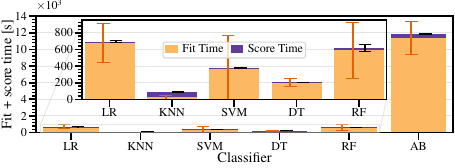}\label{fig:model_selection:times}%
  }%
  \caption{The mean of the means of the accuracy, precision, recall, and $F_1$ score ($\bar\mu_m$) over the 5-fold cross validation of all 296 scenarios as well as mean fit and scoring times. This backs our decision to use RF as a classifier. The scoring times also include scoring for \emph{balanced accuracy} and the \emph{area under the receiver operating characteristic curve (ROC AUC)} which we did not include in our final analysis for space.}%
  \label{fig:model_selection}
\end{figure}

\begin{figure*}
  \setlength{\belowcaptionskip}{-1em}
  \centering
  \subfloat[Unconstrained, P2, CoAPS, block size 1024~bytes]{
    \includegraphics{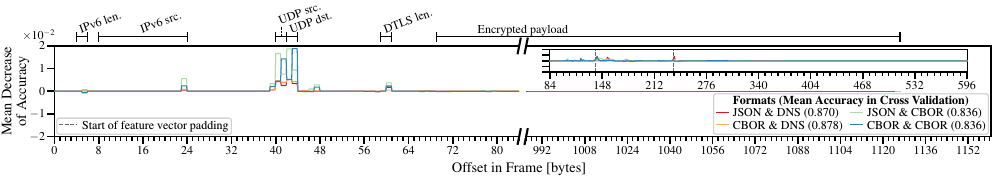}\label{fig:perm-imp-length:p2-coaps-eth-1024}%
  }\\[-0.1em]%
  \subfloat[Unconstrained, D1, OSCORE, block size 1024~bytes]{
    \includegraphics{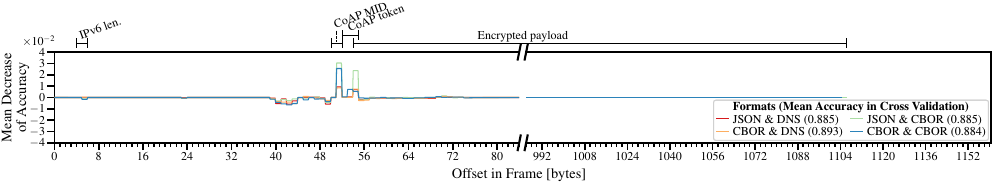}\label{fig:perm-imp-length:d1-oscore-eth-1024}%
  }
  \caption{Example results of unconstrained scenarios without block-wise transfer where length only has little permutation importance. Note the broken $x$-axes and different $y$-axis scalings.}%
  \label{fig:perm-imp-length}
\end{figure*}

\subsection{Model Selection}\label{sec:eval:model-selection}
In our machine learning analysis, we classify DNS frames as ``positive'' and data frames as ``negative'' (see the confusion matrix in \cref{fig:confusion_matrix}).
We perform 5-fold cross validation for all classifiers and determine accuracy, precision, recall, and $F_1$ score. %

In our header field analysis, we use the means $\bar\mu_m$ and $\bar\sigma_m$ of mean $\mu_{m, s}$ and standard deviation $\sigma_{m, s}$ for all metrics $m_{s, f}$ of the 296 scenarios~$s$ for each fold~$f$, $\mathbf{mean}$ and $\mathbf{std}$ being the functions to calculate the mean and standard deviation of a dataset.
\begin{align*}
  m             &\in \{\text{Accuracy}, \text{Precision}, \text{Recall}, F_1\}  \\
  s             &\in \{1, \ldots, 296\} \\
  \mu_{s, m}    &= \mathbf{mean}\left(\left\{m_{s, f}\right\}_{f = 0}^{5}\right) \\  %
  \sigma_{s, m} &= \mathbf{std} \left(\left\{m_{s, f}\right\}_{f = 0}^{5}\right) \\  %
  \bar\mu_m     &= \mathbf{mean}\left(\left\{\mu_{s, m}\right\}_{s = 1}^{296}\right)\\  %
  \bar\sigma_m  &= \mathbf{mean}\left(\left\{\sigma_{s, m}\right\}_{s = 1}^{296}\right)  %
\end{align*}

We provide these means of mean accuracy, precision, recall, and $F_1$ in \cref{fig:model_selection}.
Relatively, all classifiers perform equally well in all scenarios, \ie there is no classifier that has, \eg higher accuracy for a certain group of scenarios (not shown).
However, in 9.6\% of all scenarios, SVM only completed 2--3$\times$ of 5 folds due to memory exhaustion (not shown).
We exclude it from further consideration.
AB needs one order of magnitude more time than the other classifiers, see \cref{fig:model_selection:times}.
We exclude it as well.
Both AB and DT are not accelerated with cuML, but DT performs relatively well.
LR has the best mean accuracy and $F_1$ score, closely followed by RF (see \cref{fig:model_selection:metrics}).
RF, on the other hand, is more than one minute faster per fold on average.
DT and KNN are very comparable in their results, and while being the fastest, they perform the worst relatively when it comes to accuracy and $F_1$ score.
Additionally, DT tends to overfitting in case of large datasets, which RF solves.
In total, RF is a good compromise between best results and runtime.
Other works also use RF to identify encrypted DNS traffic~\cite{hlcw-iildot-19,cskd-pdohr-21,aaazs-rsnto-23,pammc-eiotp-24}, which bolsters our decision.
We present more detailed results of RF in Appendix~\ref{sec:rf-full}.

\section{Evaluation}\label{sec:eval}

\begin{figure*}
  \setlength{\abovecaptionskip}{0.5em}
  \setlength{\belowcaptionskip}{-1.0em}
  \centering
  \subfloat[Unconstrained, D2, Unencrypted CoAP, block size 64~bytes]{
    \includegraphics{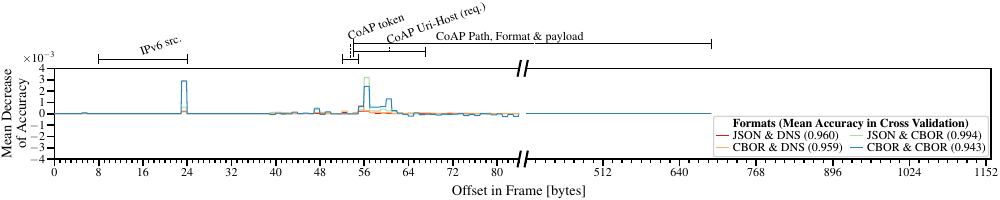}\label{fig:perm-imp-d2:coap-eth-64}%
  }\\[-0.6em]%
  \subfloat[Unconstrained, D2, Onion OSCORE, block size 1024~bytes]{
    \includegraphics{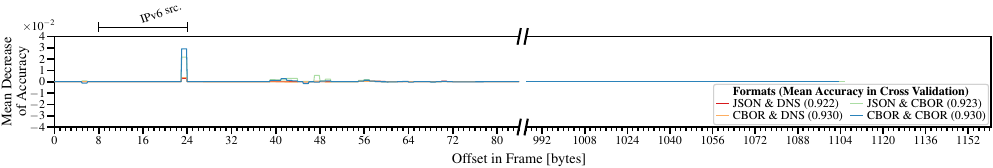}\label{fig:perm-imp-d2:onion-oscore-eth-1024}%
  }\\[-0.1em]%
  \subfloat[SCHC, \emph{min.}, D2, Onion OSCORE, block size 64~bytes]{
    \includegraphics{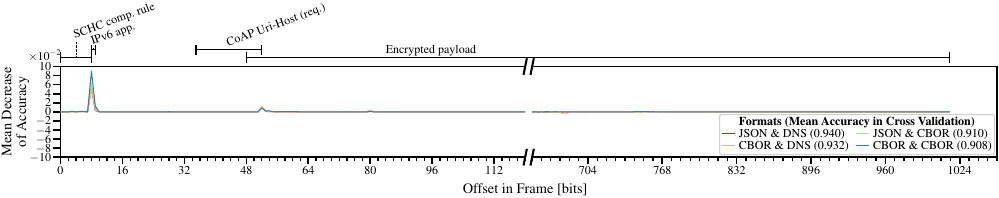}\label{fig:perm-imp-d2:onion-oscore-schc-min-64}%
  }
  \caption{Example results of permutation importances for D2 scenarios leaking source or destination information. Note the broken and differently scaled $x$-axes for unconstrained, block size, or SCHC and different $y$-axis scalings.}%
  \label{fig:perm-imp-d2}
\end{figure*}

\begin{figure*}
  \setlength{\abovecaptionskip}{0.5em}
  \setlength{\belowcaptionskip}{-1em}
  \centering
  \subfloat[Unconstrained, P1, HTTPS]{
    \includegraphics{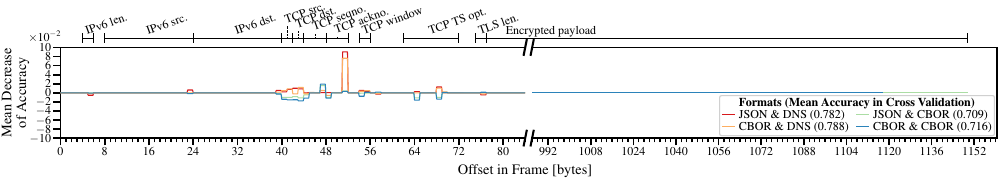}\label{fig:perm-imp-seq:p1-https}%
  }\\[-0.1em]%
  \subfloat[Unconstrained, D1, CoAPS, block size 1024~bytes]{
    \includegraphics{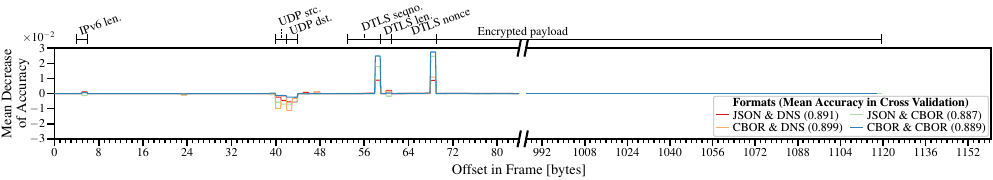}\label{fig:perm-imp-seq:d1-coaps-eth-1024}%
  }
  \caption{Example results of scenarios where monotonically growing counters had the largest permutation importances. Note the broken $x$-axes and different $y$-axis scalings.}%
  \label{fig:perm-imp-seq}
\end{figure*}

In this section, we present the results of our ML-driven analysis to distinguish non-DNS and DNS traffic and present details on the ML configuration.

\subsection{Configuration}
First, we train the RF classifier on 80\% of each dataset.
For the remaining 20\%, we test the significance of each bit or byte.
To that end, we use \emph{permutation feature importances}~\cite{sklearn-perm-imp} as mean decrease in accuracy with two permutation repeats.
We initialize both classifier and the permutation importances function with the same random seed.
This ensures that a potential change in accuracy is based on the permutation of the feature vector.
We focus on the accuracy for two reasons.
\one Our dataset is mostly balanced, and \two we account for true negatives, \ie correctly identified data frames.
The advantages of the $F_1$ score, precision, and recall do not hold in our case.
It is also worth noting that the $F_1$ score is proportional to the accuracy (see also Appendix~\ref{sec:rf-full}).

To speed up calculations in unconstrained scenarios, only the byte level can be considered.
To account for this, we modified the \emph{scikit-learn} function to permute entire 8-column groups (\ie bytes) for these scenarios, instead of using the usual single column (\ie bits).
Based on that, we map the bits or bytes to the corresponding header fields or compression residues, respectively, to get insights into the importance of a (compressed) header field.

\subsection{Finding Leaks Using Feature Importances}\label{sec:eval:leaks}

Showing all scenarios would lead to 74 plots (294 scenarios $\div$ 4 format combinations).
We present only a selection of plots that summarizes our findings.
Please find the full set of plots in our artifacts.

Each plot shows the mean decrease of accuracy for each byte or bit, highlighting the affected header fields on the top $x$-axis.
The overall mean accuracy from cross validation is noted in the legend.
We do not align the $y$-axes, \ie each $y$-axis is cut at the largest absolute change, because for permutation importance, it is more important that there \emph{is} a difference in score after shuffling a column.
For visibility reasons, we have to cut the $x$-axis.
If there are relevant changes in this part, we embed a rescaled zoom.
Exceptions where we show the full plot are scenarios with a lot of important columns containing padding of the feature vector (see \cref{sec:method:analysis}).
In any case, groups of very important padding columns are marked with a dashed line.
These correlate with the length of the frame.
We now summarize our findings.

\paragraph{Length may not be that important}
When considering the entire packet, the length of the packet is usually of low importance to distinguish frames.
We provide two of those examples in \cref{fig:perm-imp-length}.
Rather, destination information (\cref{fig:perm-imp-length:p2-coaps-eth-1024}) or monotonically growing counters, \eg non-randomized CoAP MIDs (\cref{fig:perm-imp-length:d1-oscore-eth-1024}), have higher importance.

\begin{figure*}
  \setlength{\belowcaptionskip}{-0.5em}
  \setlength{\abovecaptionskip}{0.25em}
  \centering
  \includegraphics{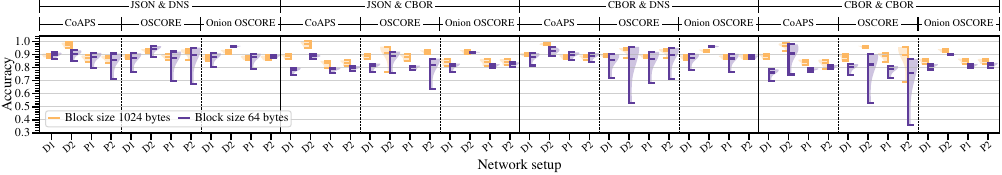}
  \caption{Violin plots of the accuracy of RF for 5-fold cross validation for block size 1024~bytes vs.~block size 64~bytes of unconstrained scenarios. Horizontal bars represent minimum, mean, and maximum of all 5 folds from bottom to top.}%
  \label{fig:blocksize-accuracy}
\end{figure*}

\begin{figure*}%
  \setlength{\belowcaptionskip}{-1em}
  \setlength{\abovecaptionskip}{0.25em}
  \centering
  \includegraphics{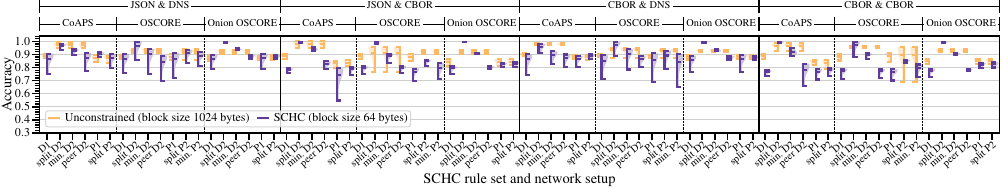}
  \caption{Violin plots of the accuracy of RF for 5-fold cross validation for unconstrained block size 1024~bytes vs.~SCHC block size 64~bytes scenarios. Horizontal bars represent minimum, mean, and maximum from bottom to top.}%
  \label{fig:link-layer-accuracy}
\end{figure*}

\paragraph{Leak 1: Destination information}
Source and destination information can have the most importance, especially when clients communicate with different DNS and data servers directly (D2).
\cref{fig:perm-imp-d2} shows some examples of bits of the source or destination address having by far the largest importance to RF, even when packets are completely unencrypted (\cref{fig:perm-imp-d2:coap-eth-64}) or when compressed to a single bit with SCHC minimum rules (\cref{fig:perm-imp-d2:onion-oscore-schc-min-64}).
In addition to addresses, unencrypted hostname information (\cref{fig:perm-imp-d2:coap-eth-64}) and transport layer ports can leak information when different ports are used for DNS and data services, even through a proxy~(\cref{fig:perm-imp-length:p2-coaps-eth-1024}).

\paragraph{Leak 2: Alternating DNS and data message patterns}
In most IoT scenarios we consider in our setup, DNS query-response pairs precede a data exchange~\cite{pammc-eiotp-24}.
This allows the classifier to recognize this pattern based on monotonically growing counters, such as TCP sequence and acknowledgment numbers, DTLS sequence numbers, CoAP MIDs, CoAP tokens, and even carelessly constructed cipher nonces.
\cref{fig:perm-imp-seq} shows a selection of scenarios where the permutation importance of these features is especially pronounced.
\cref{fig:perm-imp-length:d1-oscore-eth-1024} shows this also for the OSCORE case.
The DTLS explicit cipher nonce is leaking information (see \cref{fig:perm-imp-seq:d1-coaps-eth-1024}).
This is due to TinyDTLS, a common library we also use in our evaluation.
TinyDTLS copies the epoch (always 1 for encrypted application data) and sequence number fields for usage as cipher nonce.
This results in the same importance as the sequence number.

\subsection{Analysis of Possible Solutions}\label{sec:eval:solutions}
Beyond already known solutions, \eg binding to randomized source ports~\cite{draft-ietf-iotops-iot-dns-guidelines}, we propose the following steps to increase obfuscation of DNS traffic and close the leaks seen in \cref{sec:eval:leaks}.

\paragraph{Solution 1: Block-wise transfer}
Even though length has little importance in most scenarios, we can gain small reductions of mean accuracy when using a small block size of 64 bytes with CoAP\@.
\cref{fig:blocksize-accuracy} compares the accuracy of RF from our 5-fold cross validation as violin plot for block sizes of 1024~bytes and 64~bytes for all unconstrained and encrypted CoAP scenarios grouped by format, protocol, and network setup.
The horizontal bars show minimum, mean, and maximum from bottom to top.
For plain OSCORE in general, but also for proxy-based or split server approaches based on other protocols, we see large reductions of minimum accuracy, sometimes over~$30\%$.

\begin{figure*}
  \setlength{\belowcaptionskip}{-1em}
  \vskip-0.25em
  \centering
  \includegraphics{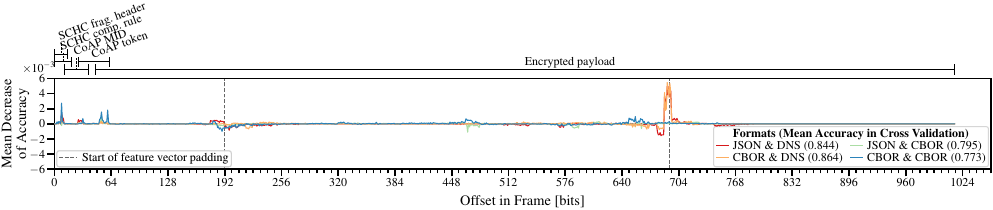}%
  \caption{Permutation importances for SCHC D2 scenario with plain OSCORE with block size 64~bytes leaking information due to its lengths.}%
  \label{fig:perm-imp-pad}
\end{figure*}

\begin{figure*}
  \centering
  \vskip-0.5em
  \setlength{\belowcaptionskip}{-1em}
  \subfloat[Plain OSCORE.]{
    \includegraphics{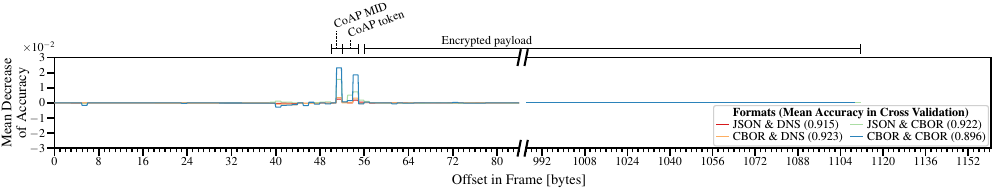}
  }\\[-0.1em]
  \subfloat[Onion OSCORE.]{
    \includegraphics{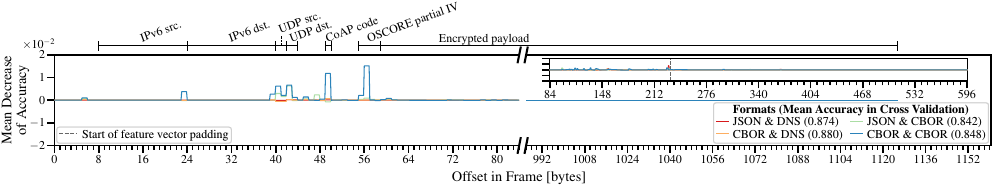}
  }
  \caption{Permutation importances for unconstrained P2 scenario with and plain OSCORE and Onion OSCORE with block size 1024~bytes. Note the broken $x$-axes. While with plain OSCORE the MID and token are the most important features, their importance fades due to randomization. Partial IV and destination information become more prominent.}%
  \label{fig:perm-imp-piv}
\end{figure*}

\paragraph{Solution 2: Peer-based SCHC rules}
Using peer-based SCHC rules, we can further reduce accuracy by obfuscating destination information and close leak 1, see \cref{sec:eval:leaks}.
\cref{fig:link-layer-accuracy} compares the accuracy of RF from our 5-fold cross validation as violin plot for unconstrained scenarios with block size 1024~bytes, \ie the same values as in \cref{fig:blocksize-accuracy}, and SCHC scenarios with 64~bytes as for all encrypted CoAP scenarios.
If there are different rule sets for SCHC, we just compare with the same unconstrained setup.
Combining block-wise transfer and SCHC, we can reduce the accuracy in most scenarios, except \emph{split} D2.
With \emph{peer} D2 we gain similar results as for P2, even with plain OSCORE, removing the necessity of the overhead of Onion OSCORE there.
However, the length, or rather the padding of the feature vector, now becomes an important feature again (see \cref{fig:perm-imp-pad}).
Nevertheless, the accuracy is only 77\%-86\% (depending on DNS and data format) compared to 91\%-100\% in unconstrained scenarios, including HTTPS (not shown, see Appendix~\ref{sec:rf-full}).
This leak comes from the last block, which is not equalized in length.

\paragraph{Solution 3: Onion OSCORE with randomized IDs via proxy}
We can reduce accuracy to a minimum when randomizing the CoAP MID and tokens as well as using Onion OSCORE via proxy compared to plain OSCORE, see \cref{fig:blocksize-accuracy,fig:link-layer-accuracy}.
This obfuscates alternating DNS and data message patterns and closes leak 2, see \cref{sec:eval:leaks}.
Looking at feature importances in \cref{fig:perm-imp-piv}, destination information but also the OSCORE Partial IV as a remaining counter help to identify DNS~frames.
Since the Partial IV is used for replay window detection, it can not be randomized.
However, there are IETF discussions to encrypt it~\cite{draft-tiloca-core-oscore-piv-enc}.
In \cref{fig:blocksize-accuracy}, we actually see an \emph{increase} in accuracy, especially in the minimum, with Onion OSCORE compared to plain OSCORE, when the \texttt{application/dns+cbor} format is used, showing again that length here becomes more important.
Similar effects are visible with SCHC in \cref{fig:link-layer-accuracy}.
In both cases, a combination of more prominent destination information (in unconstrained scenarios, ephemeral ports are used), size, and the singular Partial IV, make it easier for RF to successfully classify.
Randomizing the Partial IV in the traces to simulate its encryption reduces the accuracy to under 75\%, even 65\% in certain cases (not~shown).

\section{Related Work}\label{sec:related-work}

To the best of our knowledge, this is the first paper to analyze \one how well DoC traffic can be distinguished from CoAP data traffic and \two the potentials of header compression to omit information to further improve obfuscation.
In this section, we discuss related work in detail.

\paragraph{Privacy and security concerns related to DNS on the IoT}
Improving DNS security and privacy is an ongoing topic in Internet standardization~\cite{RFC-5452,RFC-7626}.
In particular, protection against pervasive monitoring is important since this is considered an attack~\cite{RFC-7258}.
Very recent discussions~\cite{iotops-dns-guidelines-25} about DNS security and privacy guidelines~\cite{lmcm-tosbp-25,draft-ietf-iotops-iot-dns-guidelines} for the IoT recommend additional obfuscation of header fields, such as the Server Name Indicator (SNI) in TLS with DoH, DoQ, or DoT using Encrypted Client Hello~\cite{RFC-9849}.
Our proposal to use Onion OSCORE to obfuscate hostname information within the CoAP header is in line with these recent suggestions, but has not yet been analyzed.

\paragraph{Traffic anonymization}
Eliding header fields is an anonymization technique, since removing headers does not necessarily offer privacy~\cite{s-aptp-11}.
Most approaches to anonymize traffic, such as Tor~\cite{dms-ttsgo-04}, LAP~\cite{hkpyn-lap-12}, or HORNET~\cite{cabdp-hornl-15}, are based on encryption and implement obfuscation of the sender~\cite{c-uerad-81} and ``hiding in the crowd''~\cite{rr-awtc-99} approaches.
Hiding metadata of the transport layer by encrypting transport header fields is often proposed~\cite{RFC-9065} to support privacy and to counter network ossification due to middleboxes.
Beyond privacy, RFC~8701~\cite{RFC-8701} discusses randomization of reserved header fields to counter network ossification.
All these approaches still carry (even opaque) metadata.
Completely omitting data, as we propose with SCHC, does not only improve anonymization and prevents ossification but also reduces packet sizes, which is very beneficial in the constrained IoT\@.

\paragraph{Obfuscating meta-data from DNS operators}
Oblivious DNS~\cite{semf-odns-19,draft-annee-dprive-oblivious-dns-00} provides means to decouple client information from DNS queries by encrypting query information and redirecting these queries to a special authoritative name server.
Oblivious DoH~\cite{schvv-odoh-21,RFC-9230} extended this idea for DoH by employing a dedicated DoH proxy.
Wang \etal~\cite{wkmr-pinod-21} introduced PINOT as an early alternative to that, which obfuscates source information directly on a programmable network switch.
Kulkarni~\etal~\cite{kdb-odoq-25} presented a port of Oblivious DoH for DoQ\@.
However, obfuscating the client from the recursive resolver is different challenge than this paper tries to solve.

\paragraph{Comparing CoAP to other protocols}
We compared CoAP to Information-Centric Networking (ICN) with Named Data Networks (NDN) and MQTT and found NDN to be the best performing protocol~\cite{gklp-ncmcm-18}.
This later resulted in guidance on how to deploy ICNs using CoAP~\cite{gasw-icoso-20,gasw-cosit-22}.
Nedergaard \etal~\cite{nsa-ecodhsdc-23} compared the DNS transfer protocols we use in this paper, CoAP, secured with DTLS or OSCORE, and HTTPS\@.
They found that CoAP, in both its security variants, has a performance advantage over HTTPS, even when using UDP-based HTTP/3 (over QUIC).
We compared the performance of DoC to classic DNS over UDP and DNS over DTLS~\cite{lagns-snrid-23}.
In this comparison, we included both security variants of CoAP\@.
We found that DNS over OSCORE outperforms DNS over CoAPS and DNS over DTLS\@.

\paragraph{Classifying encrypted DNS traffic and mitigation}
Wickramasinghe~\etal~\cite{wstj-sokde-25} found that most datasets used in studies that classify network traffic do not reflect current protocols (specifically TLS v1.3) and that most state-of-the-art classifiers are prone to overfitting.
They provided best practices to mitigate these issues and a new dataset.
Houser~\etal~\cite{hlcw-iildot-19} investigated the classification of DNS over TLS (DoT) traffic, including details of the queried webpage, by Random Forest (RF).
Csikor~\etal~\cite{cskd-pdohr-21} performed one of the earliest evaluations of distinguishing DoH traffic from web traffic.
They found that padding~\cite{RFC-8467} does not suffice to blend DoH traffic into web traffic, and provide extended padding techniques for DoH based on prior web traffic.
Pélissier~\etal~\cite{pammc-eiotp-24} demonstrated that even with DoH, individual IoT devices can be correctly identified with high accuracy.
Moulya and Hedge~\cite{mh-dcdoh-25} introduced a fast and lightweight classification method using RF, Catboost, and XGboost classifiers to rapidly distinguish DoH from data traffic as well as detection of malicious traffic within the DoH traffic.
They found that feature vectors based on the packet length yield the best results.
Typically, mitigations are based on EDNS{(0)} padding~\cite{RFC-7830} or random delays.
These are typically not compatible with the constrained IoT\@.
Alyami~\etal~\cite{aaazs-rsnto-23} proposed a randomized segmentation of TCP traffic as an alternative to large padding size.
However, this is not possible with CoAP block wise transfer as all blocks needs to be of the same size.

Perdisci~\etal~\cite{ppaa-ielsi-20} showed that their fingerprinting technique that is based on the domain names they query can identify IoT devices.
Building on that we identified IoT devices based on the domain names using more common classification techniques, such as Logistic Regression (LR), Support Vector Machine (SVM), or Random Forest (RF)~\cite{alabk-tbuid-25}.
In contrast to this study, both works only considered unencrypted messages and domain~names.

\section{Discussion of Limitations and Results}\label{sec:discussion}

In this section, we discuss potential limitations of our proposal and provide broader implications of our findings.

\subsection{Limitations of Approach}\label{sec:discussion:limitations}

\paragraph{Threat model}
We assume an eavesdropper is connected to the wireless domain, as wireless networks are common in IoT scenarios.
This assumption simplifies the attacker task because the attacker does not need direct infrastructure access.
However, assuming advanced attackers capable of wiretapping or accessing infrastructure devices does not contradict our observations.
The SCHC-based mitigation can be deployed on any network device between the gateway and proxy or gateway and server.

\paragraph{Method}
We focused on analyzing packet structures and using classic ML-classifiers, as temporal obfuscation techniques proposed in prior research~\cite{hlcw-iildot-19,pammc-eiotp-24,mh-dcdoh-25} are ineffective in IoT~scenarios.
This ineffectiveness arises because of two reasons.
First, low-power wireless networks with slow and random medium access, which is orders of magnitudes slower than the larger Internet, cannot cope with additional artificial delays or other common temporal obfuscation techniques~\cite{pammc-eiotp-24}.
Second, system behaviors of IoT devices require fast DNS~resolution.
After boot-up or long sleep cycles, \ie when cached results are likely expired, data must be retrieved before the device goes back to sleep.
Therefore, our proposal provides obfuscation techniques orthogonal to temporal features.

Furthermore, we used established techniques to ensure comparability with prior work.
Deep-learning techniques will likely yield higher accuracy when distinguishing DNS traffic, but conflict with this objective.

\paragraph{Dataset, including usage of HTTP Archive}
To the best of our knowledge, there is currently no freely available, large-scale dataset for constrained IoT traffic that includes both structured data objects and resource locators (URLs) referencing these objects.
When collecting data from the HTTP Archive~\cite{httparchive-bigquery}, we aimed for a tradeoff between the availability of Internet-based M2M communication data (JSON objects) and a reasonable alignment to the constrained IoT.
While the HTTP Archive only contains Web requests and responses, these also include many JSON-based REST API calls.
Even though not a perfect match, they fit the M2M communication patterns of CoAP when accounting for the smaller IoT packet sizes.
While largely synthesized from real-world data, we believe our data corpus can serve as valuable input for future studies on DNS privacy.
Features such as packet lengths can be leveraged in unconstrained scenarios, while the SCHC and block-wise parts offer insights into compressed and fragmented IoT traffic, in which these features are omitted or adjusted.
The Onion OSCORE part offers insights into IoT traffic, which either obfuscates or randomizes many features.
We note the caveat that URI paths of data objects are longer than in typical CoAP scenarios since they are inherited from HTTP traffic.
We do not recommend this dataset for isolated DoH studies, though, because we intentionally weakened encryption using TLS~1.2 for better comparison of HTTPS and CoAP\@.

\subsection{Discussion of Results}\label{sec:discussion:results}

\paragraphS{How should our results be interpreted}
Our techniques do not minimize distinction accuracy, nor was this our objective. Our aim was to demonstrate and quantify the (notable) effect of common standard IoT protocols on lowering the detection accuracy. Combining with other, already established obfuscation techniques may help to reduce distinction further.
The lowest achievable accuracy is hard to determine and depends on many factors, such as individual costs or gain~\cite{bg-cecav-13}. 
50\% accuracy, which represent an `attack' of randomly guessing the type of packet, is typically understood as the lowest accuracy necessary to render an attack useless~\cite{llr-miadc-21}.
Current results are much larger than that.
Our results for DoH also show higher accuracy than prior work.
We did not analyze this in detail but it may easily be explained by \one we use weaker encryption with TLS-PSK and TLS v1.2 to stay comparable to CoAPS (see \cref{sec:method:network}) and \two we look at the full packet as feature vector, instead of more scalar features, providing more potential for ML-classifiers to distinguish packets (see \cref{sec:method:analysis:vec}) in exchange for slower computation (\cref{sec:eval:model-selection} shows training times of mean 10 minutes per dataset).

\paragraphS{Can SCHC header compression and CoAP block-wise transfer decrease distinction of DNS from data traffic}
Yes.
Replacing information with an opaque rule ID using peer-based rule sets decreases the capability of a classifier to distinguish encrypted DNS and data traffic.
In SCHC scenarios, we observe lower accuracy than in unconstrained scenarios throughout our header field analysis.
However, in block-wise transfer scenarios, the last block often leaks enough information to keep the accuracy above 50\%.
This could be circumvented by using the padding option proposed in~\cite{draft-ietf-core-cacheable-oscore} to extend only the last block to full block length.
Compared to using EDNS{(0)} padding, this has the additional advantage that data packets can be padded, which improves obfuscation.
As the block size is a fixed known length, prior knowledge of data traffic, \eg as proposed by Csikor~\etal~\cite{cskd-pdohr-21}, is not required for a padding strategy. This reduces complexity, which is particularly important for constrained devices.
Like EDNS{(0)} padding, our solutions are based on pre-established or proposed IETF standards~\cite{RFC-9363,draft-toutain-schc-coreconf-management,draft-ietf-schc-universal-option,draft-ietf-core-comi,RFC-7959,draft-ietf-core-cacheable-oscore} and merely require reconfiguration, or in case of the CoAP padding option, small updates.
SCHC rules, including peer identities, can either be directly shipped with the firmware updates or updated via a YANG data model~\cite{RFC-9363}.
With CORECONF~\cite{draft-ietf-core-comi,draft-toutain-schc-coreconf-management,draft-ietf-schc-universal-option}, we can even reuse the same CoAP-based obfuscation techniques (\ie encrypted block-wise transfer).
Placing the SCHC compressor/decompressor closer to the proxy or servers extends our SCHC solution outside the wireless domain.
If two SCHC hops know about each other, \eg using CORECONF or more generally NETCONF~\cite{RFC-4741}, our proposal can also be deployed beyond the constrained domain.

\paragraphS{Should header data be obscured}
With the introduction of QUIC and Encrypted Client Hello in TLS, the practicality of obfuscating transport header information has been widely discussed.
In constrained IoT scenarios, the main concerns relate \one to message sizes, \two ossifying fields to enable compression, or \three endpoints being unaware of the constrained link~\cite{RFC-9065}.
We argue for not sending certain network, transport, or application header fields at all.
Since we propose the usage of SCHC, a technique established for constrained networks, a negative effect on message size and compression is not expected.
SCHC provides fragmentation if needed, and compression can be used for obfuscation.

\paragraphS{How does \texttt{application/dns+cbor} encoding impact the distinction of DNS}
Throughout our evaluation, using \texttt{application/dns+cbor} yields either comparable or better results than using the classic DNS message format, especially when using CBOR as a data format because it aligns lengths.
As such, depending on the use case and network setup, we recommend \texttt{application/dns+cbor} to make DNS and data traffic harder to distinguish.
More improvement is provided by CoAP block-wise transfer (especially if the block option is protected by CoAPS or OSCORE), which aligns the lengths of frames to a fixed size.

\paragraphS{Which obfuscation techniques should I use}
Encryption, equaling out packet lengths, and header elision together provide a solid foundation for obfuscation.
Peer-based SCHC rules reduces the need for a centralized proxy to increase obfuscation of traffic patterns when link layer addresses are randomized.
These peers do not need to be multiple servers.
Sharing the load over \emph{multiple} proxies is also viable, \eg to use caching advantages of proxies. %
When using peer-based SCHC rules, the overhead of Onion OSCORE also becomes unnecessary, as relevant OSCORE information can be obfuscated by header elision.
An extension of DoC for Oblivious DNS~\cite{semf-odns-19} to anonymize clients from recursive resolvers would help to reduce the risk of Internet~centralization even further.

\paragraphS{Does DoC provide better privacy than DoH}
With RF we observe lower accuracy for DoH than for DoC.
Our header field analysis points to the additional affirming counter in TinyDTLS, which uses the record epoch and sequence number as explicit nonce.
Apart from this specific implementation behavior, DoC over OSCORE together with SCHC can obfuscate DNS traffic in data traffic even better than our weakened DoH setup can.
We found that randomized counters, in case of CoAP MID and using tokens, and the obfuscation of OSCORE Partial~IV, as proposed in~\cite{draft-tiloca-core-oscore-piv-enc}, can significantly contribute to this advantage.
These mechanisms cannot easily be copied to DoH when using HTTP/2, because order-preserving sequence numbers in TCP cannot be randomized or obfuscated.
In DoH, HTTP/3 would be needed where the packet numbers of QUIC and their lengths are also protected by encryption~\cite{RFC-9000,RFC-9001}.

\section{Conclusion}\label{sec:conclusion}

This paper closes a gap by analyzing DNS over CoAP (DoC) as an obfuscation technique for DNS\@.
Prior work only investigated DNS over TLS (DoT), DNS over HTTPS (DoH), or DNS over QUIC (DoQ).
We provide a data corpus that provides potential insights for future studies into DoC\@.
Using this data corpus we utilize permutation importance for Random Forest to investigate which header leaks information.
We show that DNS over CoAP (DoC) can provide good obfuscation and provide guidelines how to improve it even further.
These include using Object Security for Constrained RESTful Environments (OSCORE)~\cite{RFC-8613} with randomized identifiers such as message IDs and tokens and block-wise transfer.
We can even improve on that by using the constrained use case as an advantage and omit leaking information altogether using Static Context Header Compression (SCHC).
When using dedicated header compression rule sets for the upstream DNS and data servers we get their accuracy down to 77\%, similar to a proxy-based approach, closing identified leaks.

Our data corpus has potential for extension which will inform future work.
Once DTLS 1.3 implementations reach suitable stability with CoAP, the data corpus should be amended with such traces.
Implementation of encrypted OSCORE Partial IV~\cite{draft-tiloca-core-oscore-piv-enc} or the CoAP padding option~\cite{draft-ietf-core-cacheable-oscore} will also provide material for amendment.
The advent of Oblivious DNS provides potential of an analysis how an oblivious version of DoC performs.
Real world traces of DoC, especially when mixing with or even being used for classic Internet deployments, will provide valuable insights into the privacy enhancements of DoC once it sees wider use.

\paragraph{Artifacts}
All artifacts of this work, including the data corpus of traffic traces, implementations, and experiment results, as well the complete plots of our header field analysis are available under \textbf{\href{https://doi.org/10.5281/zenodo.19697848}{doi:10.5281/zenodo.19697848}}.

\section*{Acknowledgment}

We would like to thank the anonymous shepherd and reviewers for their valuable feedback on this paper.
Special thanks to Christian Amsüss for his help and guidance in beating \emph{aiocoap} into shape for the evaluation of this paper and Mikolai Gütschow for reviewing the artifacts.
This work was supported in parts by the German Federal Ministry of Research, Technology and Space (BMFTR) within the research project C-ray4edge (grants 16KIS1694K and 16KIS1695).

\label{lastpage} %
\bibliographystyle{IEEEtran}
\bibliography{./local,rfcs,hypermedia,ids,internet,iot,own,security}

\appendices

\section{Random Forest Results for All Scenarios}\label{sec:rf-full}

\begin{figure}
  \setlength{\abovecaptionskip}{5pt plus 3pt minus 2pt}
  \includegraphics{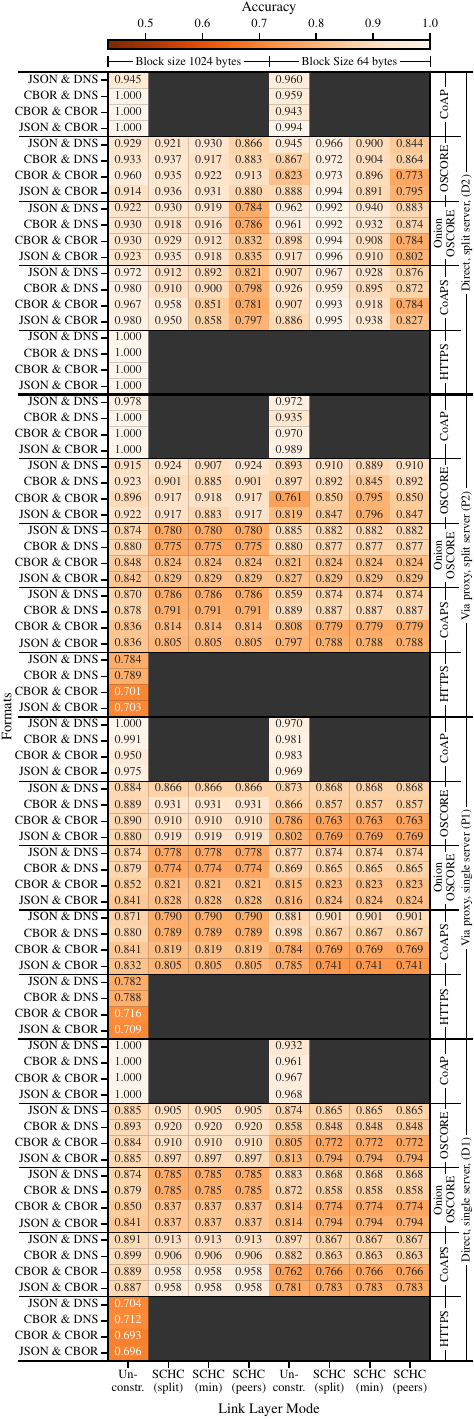}
  \caption{Mean accuracy for 5-fold cross validation of Random Forest. Dark gray cells mean that the scenario was not tested by design.}%
  \label{fig:binvec_rf_accuracy}
\end{figure}

\begin{figure}
  \setlength{\abovecaptionskip}{5pt plus 3pt minus 2pt}
  \includegraphics{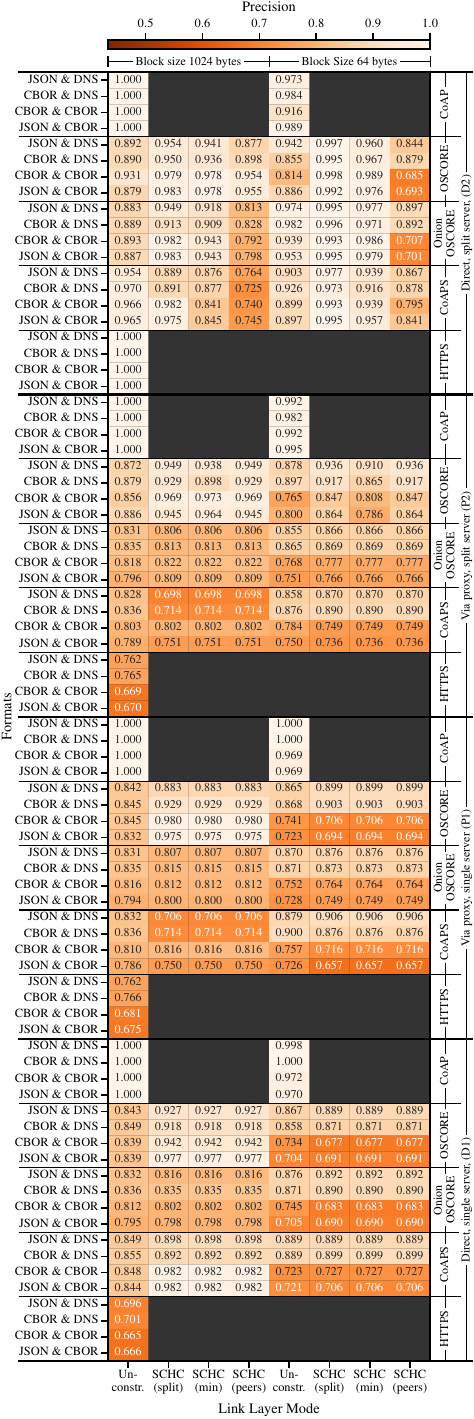}
  \caption{Mean precision for 5-fold cross validation of Random Forest. Dark gray cells mean that the scenario was not tested by design.}%
  \label{fig:binvec_rf_precision}
\end{figure}

\begin{figure}
  \setlength{\abovecaptionskip}{5pt plus 3pt minus 2pt}
  \includegraphics{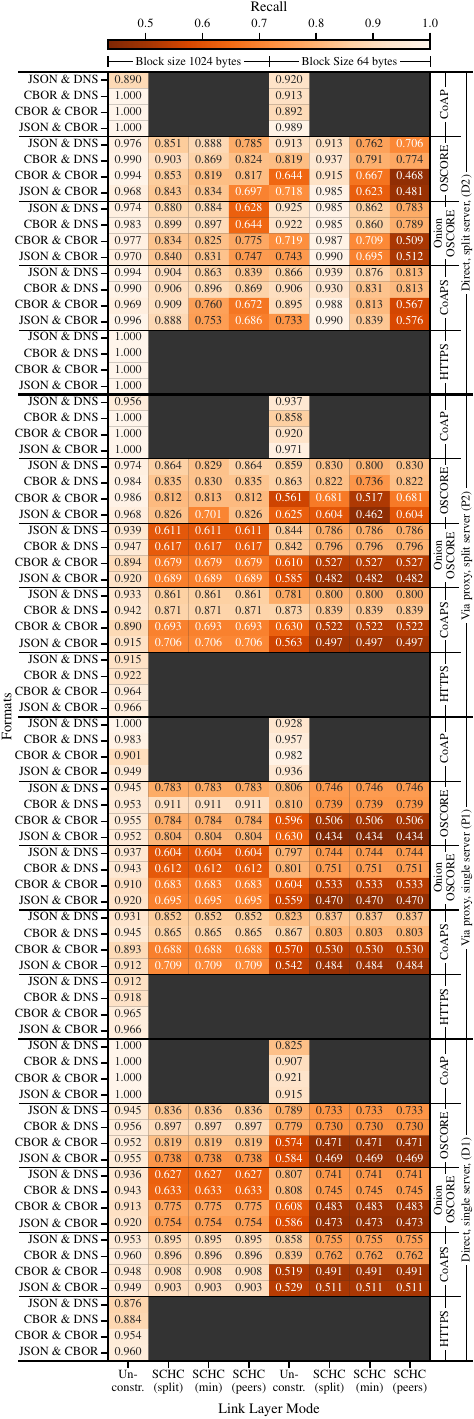}
  \caption{Mean recall for 5-fold cross validation of Random Forest. Dark gray cells mean that the scenario was not tested by design.}%
  \label{fig:binvec_rf_recall}
\end{figure}

\begin{figure}
  \setlength{\abovecaptionskip}{5pt plus 3pt minus 2pt}
  \includegraphics{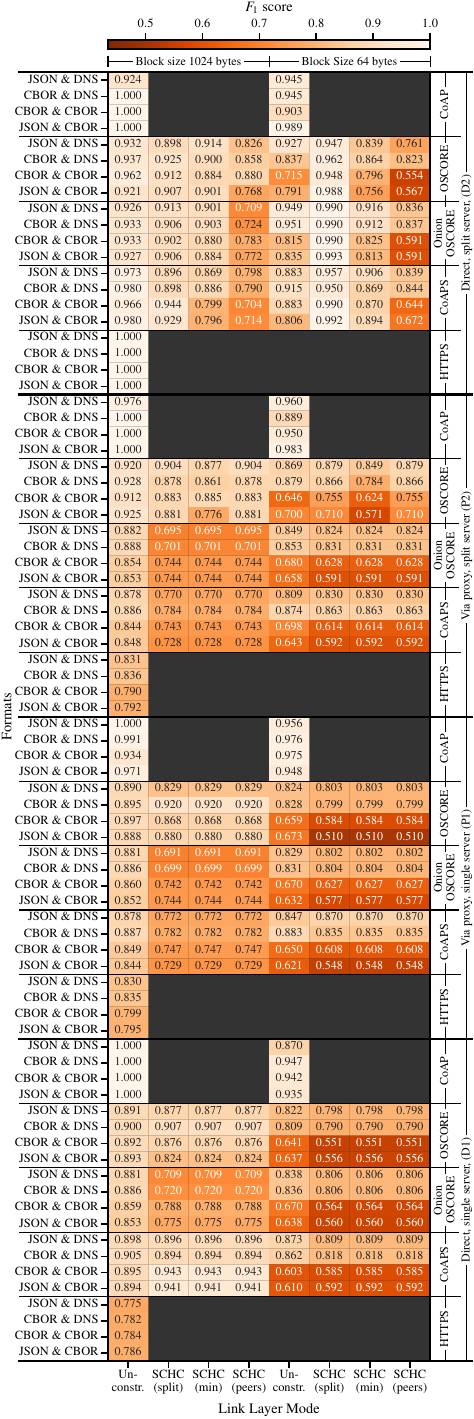}
  \caption{Mean $F_1$ score for 5-fold cross validation of Random Forest. Dark gray cells mean that the scenario was not tested by design.}%
  \label{fig:binvec_rf_f1_score}
\end{figure}

\cref{fig:binvec_rf_accuracy,fig:binvec_rf_precision,fig:binvec_rf_recall,fig:binvec_rf_f1_score} show the mean accuracy, precision, recall, and $F_1$ score for 5-fold cross validation of Random Forest as annotated heatmaps.
The $x$-axes summarize the link layer mode (unconstrained or SCHC and which rule set was used with SCHC) and the block size.
The $y$-axes summarize network architecture (D1, P1, P2, D2), protocol (HTTPS, CoAPS, Onion OSCORE, CoAP) and format combination (JSON \& CBOR, CBOR \& CBOR, CBOR \& DNS, JSON \& DNS), roughly ordered by metric and group.

\clearpage

\end{document}